\documentclass[a4paper,12pt]{article}
\usepackage{latexsym,amssymb}
\usepackage{amsmath,amssymb,amsxtra,amscd,amsthm}
\usepackage[mathscr]{eucal}
\DeclareFontFamily{U}{rsfs}{\skewchar\font127 }
\DeclareFontShape{U}{rsfs}{m}{n}{
   <5> rsfs5
   <6> rsfs6
   <7> rsfs7
   <8> rsfs8
   <9> rsfs9
   <10> rsfs10
   <10.95> rsfs11
   <12> rsfs12
   <14.4> rsfs14
   <17.28> rsfs17
   <20.74> rsfs20
   <24.88> rsfs25
   <29.86-> rsfs30}{}
\DeclareMathAlphabet\scr{U}{rsfs}{m}{n}

\def\mZ{\mathbb{Z}}

\def\tL{\mathrm{L}}

\def\tS{\mathrm{S}}

\newfont{\HUGE}{cmssbx10 scaled 4000}

\def\bb1{\textup{\small{1}} \kern-3.8pt \textup{1}}

\def\SL2Z{\tS\tL(2,\mZ)}

\numberwithin{equation}{section}

\providecommand{\href}[2]{#2}

\newtheoremstyle{plain} 
  {5pt}
  {10pt}
  {\rmfamily} 
  {}
  {\scshape} 
  {}
  {\newline} 
  {}
\swapnumbers
\theoremstyle{break}

\def\a{\alpha}

\def\g{\gamma}
\def\l{\lambda}

\newcommand{\OSp}{\mathop{\rm {}OSp}}

\newcommand{\so}{\mathfrak{so}}

\newcommand{\sym}{\mathfrak{sp}}

\newcommand{\osp}{\mathfrak{osp}}

\newcommand{\ft}[2]{{\textstyle\frac{#1}{#2}}}

\begin{document}
\begin{titlepage}
\begin{flushright}
DISTA-2008\\
hep-th/yymmnnn
\end{flushright}
\vskip 1.5cm
\begin{center}
{\LARGE \bf
Pure Spinor Formalism for \\ $\mathrm{Osp}(\mathcal{N}|4)$ backgrounds  }
\vfill
{\large Pietro Fr\'e$^{~a\,}$\footnote{fre@to.infn.it} and Pietro Antonio Grassi$^{~b\,}$\footnote{pgrassi@cern.ch}} \\
\vfill {
$^{a}$
Department of Theoretical Physics, University of Torino, v. Giura 1, \\
10100 Torino, Italy and INFN - Torino, Italy,
\vskip .2cm
$^{b}$
 DISTA, University of Eastern Piedmont, v. Bellini 25/g, \\
 15100 Alessandria, Italy, and INFN - Torino, Italy.
}
\end{center}
\vfill
\begin{abstract}
We start from the Maurer-Cartan (MC) equations of the $\mathrm{Osp}(\mathcal{N}|4)$ superalgebras satisfied by the left-invariant
super-forms realized on supercoset manifolds of the corresponding
supergroups and we derive  some new pure spinor constraints. They
are obtained by "ghostifying" the MC forms and extending the differential $d$ to a BRST
differential. From the superalgebras  $\widehat{\mathbb{G}} = \mathrm{Osp}(\mathcal{N}|4)$  we single out different subalgebras
$\mathbb{H} \subset \widehat{\mathbb{G}}$  associated with the different cosets
$\mathrm{\widehat{G}/H}$: each choice of $\mathbb{H}$ leads to a
different weakening of  the pure
spinor constraints. In each case, the number of parameter is counted and we show that in the
cases of $\mathrm{Osp}(6|4)/ \mathrm{U(3)}\times \mathrm{SO(1,3)}$, $\mathrm{Osp}(4|4)/ \mathrm{SO(3)} \times \mathrm{SO(1,3)}$ and
finally $\mathrm{Osp}(4|4)/ \mathrm{U(2)} \times \mathrm{SO(1,3)}$ the bosonic and fermionic degrees of freedom
match in order to provide a $c=0$ superconformal field theory. We construct both the Green-Schwarz and the pure spinor sigma model
for the case $\mathrm{Osp}(6|4)/ \mathrm{U(3)}\times \mathrm{SO(1,3)}$ corresponding to $\mathrm{AdS_4} \times \mathbb{P}^3$.
The pure spinor sigma model can be consistently quantized.

\end{abstract}
\vfill
\vfill

\date{February 2008}
\end{titlepage}

\tableofcontents

\section{Introduction}

Due to the recent developments
in constructing the AdS/CFT dual of supeconformal Chern-Simons theory \cite{Schwarz:2004yj,Bagger:2006sk,Aharony:2008ug},
it became rather important to develop
a formalism to quantize  superstrings on backgrounds of the form $\mathrm{AdS_4} \times \mathbb{P}^3$
\cite{Nilsson:1984bj} (see also \cite{Sorokin:1985ap}).
The formalism suitable for that purpose is, of course, the pure spinor formalism \cite{Berkovits:2000fe}
where the supersymmetry and the bosonic isometries of the target space can be maintained manifest to all stages of the computations.
In addition, due to the manifest supersymmetry, the coupling with
the RR fields is simplified, or to be more precise, they appear linearly coupled to the sigma
model fields \cite{BerkovitsADS}. Naturally, the RR fields appear also in the higher-component expansion of the
superfields entering the Green-Schwarz sigma models (see for example \cite{Metsaev:1998it}), but
the pure spinor sigma model contains a new coupling with the RR wich breaks the kappa-symmetry
of the action Green Schwarz action leading to a quantizable field theory model \cite{Berkovits:2004xu}.
\par
In the present paper, we first recall some of the ingredients of the construction, but
differently from the construction performed in  paper \cite{marp},
we observe that the number of the pure spinor degrees of freedom for different anti-de Sitter
compactifications can be directly obtained by analyzing just the Maurer-Cartan forms related with their cosets.
Given the supergroup $\mathrm{Osp}(\mathcal{N}|4)$, we construct the corresponding Maurer-Cartan equations
(see \cite{Castellani:1991et} for a complete description of these group manifold and the explicit form
of the MC forms). This is a standard procedure. Next we ghostify the Maurer Cartan system   extending the superforms to generalized forms
by shifting each of  fermionic forms by means of a commuting $0$-form denoted in the text by $\Lambda$ (see also \cite{Fre:2008qw}).
In addition, we extend the differential $\mathrm{d}$ with a BRST differential $\mathcal{S}$. The latter, is nilpotent
only upon some constrains on $\Lambda$'s. As was explained in \cite{marp}, projecting the BRST variations of the
target space fields onto the worldsheet and by identifying the commuting $0$-forms $\Lambda$ with the pure spinor
on the worldsheet, we find some new constraints for the pure spinor fields \cite{Fre:2008qr}\footnote{In 10d, in
\cite{Berkovits:2000fe} the Cartan pure spinors are taken into account \cite{Howe:1991bx}. They look different, but as was discussed in
\cite{Fre:2008qr} they coincide
upon some redefinitions.}. As shown in \cite{Fre:2008qr}, the new pure spinor constraints are equivalent to the original ones and therefore,
we obtain a new form of the sigma model action which has the same coupling as those in the
work of Berkovits and Howe \cite{Berkovits:2001ue}.
\par
Since our approach is  meant to work for any background, we can apply it to the cases with less conserved supersymmetry as
the background $\mathrm{AdS_4} \times \mathbb{P}^3$. However, before getting to
this particularly relevant example, we analyze several different ways to produce consistent backgrounds for critical and
non-critical dimensions by modding the supergroup $ \widehat{\mathrm{G}} \equiv \mathrm{Osp}(\mathcal{N}|4)$,
with respect to different subgroups $ H \subset \widehat{\mathrm{G}}$ which are always chosen bosonic.
In particular we consider the supergroup manifold $\mathrm{Osp}(\mathcal{N}|4)$.
This case does not lead to any consistent background since the solution of the pure spinor constraints has zero non-vanishing components.
Then, we move to the case of  $\mathrm{Osp}(\mathcal{N}|4)/ \mathrm{SO}(\mathcal{N} -1) \times \mathrm{SO(1,3)}$. Now, since
we have modded out the subgroup $\mathrm{SO}(\mathcal{N} -1) \times\mathrm{ SO(1,3)}$, we have to consider the nilpotency of the BRST differential
modulo the gauge symmetry of the subgroup. This leads to
new pure spinor constraints. We found that the matching between bosonic and fermionic degrees of
freedom is possible only for $\mathcal{N} =4$ and the bosonic subset of the coset
$\mathrm{Osp}(4|4)/ \mathrm{SO(3)} \times \mathrm{SO(1,3)}$ corresponds to $\mathrm{AdS_4} \times \mathrm{S^3}$.
So, it would be a consistent background for a 7 dimensional supergravity. We do not dwell on this case in the present paper.
\par
We move to the more interesting example where the subgroup is $U(\mathcal{N}/2) \times \mathrm{SO(1,3)}$. There
we find a new modified forms of the pure spinor constraints which we are able to solve.
We found that there are two cases where the matching to the bosonic and fermionic degrees of freedom takes
place, namely for $\mathcal{N} =4$ and $\mathcal{N}=6$.
The bosonic part of these cosets correspond to the backgrounds $\mathrm{AdS_4} \times \mathbb{P}^1$ and $\mathrm{AdS_4} \times \mathbb{P}^3$.
They both have RR fields in the spectrum, in particular for the first case there is a two form in $\mathbb{P}^1$ which coincides
with the K\"ahler form on $\mathbb{P}^1$ and with a RR 4-form on $\mathrm{AdS_4}$.
The same for the case of $\mathbb{P}^3$.
The first background would be a consistent  background for non-critical string in 6 dimensions and it might
be verified that that solution exists for supergravity in d=6 with
$N=4$ supersymmetry (corresponding to 16 supercharges in 4 dimensions).
The second example is of course more interesting and it has $\mathcal{N}=6$ supersymmetry.
\par
The last example is a critical theory in 10 dimensions and therefore we can write down the corresponding sigma model.
This is done in the last section where all the ingredients are described and the action is also presented.
In addition, it has been noticed that  by decomposing the MC forms into $\mathrm{SO(1,3)}$ representations,
one finds that the superalgebra admits the famous $\mathbb{Z}_4$ discrete symmetry.
The action is constructed  respecting such a symmetry. We start by constructing the Green-Schwarz action with $\kappa$-symmetry. The action
takes the standard form of a quadratic action where the principal term is the usual quadratic action written of the bosonic MC forms; the
second addend contains the WZ terms which can also be written as a quadratic expression in the fermionic MC forms. This is
a normal evenience
for backgrounds of the form $\mathrm{AdS_q} \times \mathbb{S}^p$ \cite{Berkovits:1999zq}.
It can be shown that   $\kappa$-symmetry reduces correctly the
24 fermions to the 16 light-cone degrees of freedom and that reparametrization invariance reduces the bosonic coordinates to light-cone ones.
\par
While completing  the present paper, two other contributions \cite{Arutyunov:2008if} and \cite{Stefanski:2008ik} appeared on
arXive with a partial overlap with our results. We therefore do not discuss $\kappa$-symmetry, but we proceed with the construction
of the pure spinor sigma model. The resulting sigma model has 24 manifest supersymmetries and it can be covariantly quantized.
In addition, since the formalism to construct the pure spinor sigma model given a Green-Schwaz action was
discussed in several papers, we refer to \cite{Adam:2007ws} since it is adapted also
to non-critical backgrounds with less supersymmetry \cite{Adam:2006bt}.
\par
There are some important remarks that we would like to make: first, the pure spinor sigma model seems
to respect, at least at the first expansion in $\a'$, the cancellation between bosonic and fermionic degrees of freedom.
Indeed the 10 dimensional bosonic coordinates are cancelled by the 24 fermionic coordinates and by the
14 pure spinor fields and their conjugated. In order to compare it with the most studied case of $\mathrm{AdS_5} \times \mathbb{S}^5$, we
recall that since there are 32 manifest supersymmetries we need to have 22 pure spinor fields in order to
saturate the central charge. In \cite{BerkovitsADS}, it has been discussed the pure spinor constraints for closed type IIB superstrings
(see also \cite{Berkovits:2007rj} for pure spinor constraints written in $\mathrm{PSU(2,2|4})$ basis) and it has been noticed
that they are sufficient to compensate the rest of the coordinates. In the case of $\mathrm{AdS_4} \times \mathbb{P}^3$, with less
conserved supersymmetry we consistently remove 8 fermionic coordinates and 8 pure spinors from the 32 fermionic $\theta$ coordinates and from
the 22 pure spinors, leading to the result of the present paper. It can be also checked that the pure spinor constraints derived
as in \cite{marp} (the complete discussion will be presented elsewhere \cite{marp2}) lead to the same conclusion. Not only that. In
the forthcoming paper \cite{marp2} we show that the pure spinor action and the BRST
transformation rules derived here from the algebraic structure of the Maurer Cartan
system can be obtained systematically by localizing on the chosen
supergravity background $\mathrm{AdS_4} \times \mathbb{P}^3$ the
general action discussed in \cite{marp}. Secondly we note that the
nominator
supergroup $\widehat{\mathrm{G}}=\mathrm{Osp(6|4)}$ in the supercoset is a super-Calabi-Yau and therefore, it is conceivable that the cancellations
between bosonic and fermionic degrees of freedom happen also here as   $\mathrm{PSU}(2,2|4)$. However, the proof of the
conformal invariance given in \cite{Berkovits:2004xu} does not seem to be possible using the technique described in
\cite{Berkovits:1999im}. Thirdly, the construction of non-local charges, and the analysis of the integrability can be extended to quantum
level as in \cite{Berkovits:2004jw}.

\section{The $\OSp (\mathcal{N}|4)$ supergroup, its superalgebra and its supercosets}
\label{OSPsection}

\subsection{The superalgebra}
The real form $\osp(\mathcal{N}|4)$ of the complex  $\osp(\mathcal{N}|4,\mathbb{C})$
Lie superalgebra which is relevant for the study
of $\mathrm{AdS_4} \times \mathcal{G}/\mathcal{H}$
compactifications is that one where the ordinary Lie subalgebra is the following:
\begin{equation}
\sym(4,\mathbb{R}) \, \times \, \so(\mathcal{N})\, \subset \, \osp(\mathcal{N}|4)
\label{realsuba}
\end{equation}
This is quite obvious because of the isomorphism $\sym(4,\mathbb{R}) \, \simeq
\, \so(2,3)$ which identifies $\sym(4,\mathbb{R})$ with the isometry algebra
of anti de Sitter space. The compact algebra $\so(\mathcal{N})$ is instead the
R-symmetry algebra acting on the supersymmetry charges.
\par
The superalgebra $\osp(\mathcal{N}|4)$ can be introduced as follows: consider the two graded $(4+\mathcal{N}) \times (4+\mathcal{N})$ matrices:
\begin{equation}
  \begin{array}{ccccccc}
    \widehat{C} & = & \left( \begin{array}{c|c}
      C\, \gamma_5 & 0 \\
      \hline
      0 & - \frac{\rm i}{4\, e} \, \mathbf{1}_{\mathcal{N}\times \mathcal{N}} \\
    \end{array}\right)  & ; & \widehat{H} & = &\left( \begin{array}{c|c}
      {\rm i} \, \gamma_0\,\gamma_5 & 0 \\
      \hline
      0 & \, - \, \frac{1}{4\, e} \, \mathbf{1}_{\mathcal{N}\times \mathcal{N}} \\
    \end{array} \right) \\
  \end{array}
\label{OmandHmat}
\end{equation}
where $C$ is the charge conjugation matrix in $D=4$. The matrix
$\widehat{C}$ has the property that its upper block is
antisymmetric while its lower one is symmetric. On the other hand, the
matrix $\widehat{H}$ has the property that both its upper and lower blocks are
hermitian. The $\osp(\mathcal{N}|4)$ Lie algebra is then
defined as the set of graded matrices $\Lambda$ satisfying the two
conditions:
\begin{eqnarray}
  \Lambda^T \, \widehat{C} \, + \, \widehat{C} \, \Lambda &
  = & 0 \label{ortosymp}\\
\Lambda^\dagger \, \widehat{H} \, + \, \widehat{H} \, \Lambda &
  = & 0 \label{realsect}
\end{eqnarray}
Eq.(\ref{ortosymp}) defines the complex $\mathrm{osp(\mathcal{N}|4)}$
superalgebra while eq.(\ref{realsect}) restricts it to the
appropriate real section where the ordinary Lie subalgebra is
(\ref{realsuba}).
 The specific form of the matrices $\widehat{C}$ and
$\widehat{H}$ is chosen in such a way that the complete solution of
the constraints (\ref{ortosymp},\ref{realsect}) takes the following
form:
\begin{equation}
  \Lambda \, = \, \left( \begin{array}{c|c}
   -  \ft 14 \, \omega^{ab} \, \gamma_{ab} \, - \,2\, e \, \gamma_a \, \gamma_5 \, E^a \ & \psi_A \\
   \hline
    4 \, {\rm i} \, e \, \overline{\psi}_B \, \gamma_5 & - \, e \,\mathcal{A}_{AB} \
  \end{array} \right)
\label{lambamatra}
\end{equation}
and the Maurer-Cartan equations
\begin{equation}
  d \, \Lambda \, + \, \Lambda \, \wedge \, \Lambda \, = \, 0
\label{Maurocartanosp1}
\end{equation}
read as follows:
\begin{eqnarray}
d \omega^{ab} - \omega^{ac} \, \wedge \,\omega^{db} \, \eta_{cd} + 16 e^2 E^a \,
\wedge \,E^b &=&
-{\rm i}\,  2 e \,  \overline{\psi}_A \, \wedge \gamma^{ab} \gamma^5 \psi_A, \nonumber \\
d E^a - \omega^a_{\phantom{a}c}\, \wedge \, E^c &=& {\rm i} \ft 12 \, \overline{\psi}_A \, \wedge \,\gamma^a
\psi_A, \nonumber\\
d \psi_A - \frac{1}{4} \omega^{ab}\, \wedge \, \gamma_{ab} \psi_A
- e {\mathcal{A}}_{AB} \, \wedge \,\psi_B &=&  2 e \,
 E^a \, \wedge \,\gamma_a \gamma_5 \psi_A, \nonumber\\
d {\mathcal{A}}_{AB} - e  {\mathcal{A}}_{AC}\, \wedge \, {\mathcal{A}}_{CB} &=& 4 \, {\rm i}  \overline{\psi}_A \,
\wedge \,\gamma_5 \psi_B\,.
\label{orfan25}
\end{eqnarray}
Interpreting $E^a$  as the vielbein, $\omega^{ab}$ as the spin connection, and
$\psi^a$ as the gravitino $1$-form, eq.s (\ref{orfan25}) can be
viewed as the structural equations of a supermanifold $\mathrm{AdS}_{4|\mathcal{N}\times 4}$
extending anti de Sitter space with $\mathcal{N}$ Majorana
supersymmetries. Indeed the gravitino $1$--form is a Majorana spinor since, by construction, it satisfies the reality condition
\begin{equation}
  C \, \overline{\psi}_A^T \, = \, \psi_A\,, \quad \quad \overline{\psi}_A \,
  \equiv
\, \psi_A^\dagger \, \gamma_0\,.
\label{majorancondo}
\end{equation}
The supermanifold $\mathrm{AdS}_{4|\mathcal{N}\times 4}$ can be identified with the following supercoset:
\begin{eqnarray}
 \mathcal{M}^{4|4\mathcal{N}}_{osp} & \equiv & \frac{\mathrm{Osp(\mathcal{N} \, | \, 4
})}{\mathrm{SO(\mathcal{N})} \times \mathrm{SO(1,3)}}
\label{ospsuper4_4n}
\end{eqnarray}
Alternatively, the Maurer Cartan equations
can be written in the following more compact form:
\begin{eqnarray}
d \Delta^{xy} + \Delta^{xz} \, \wedge \,\Delta^{ty} \, \epsilon_{zt} &=&
 -\, 4 \, {\rm i}\,  e \,  {\Phi}_A^x \, \wedge \, {\Phi}_A^y, \nonumber \\
d {\mathcal{A}}_{AB} - e  {\mathcal{A}}_{AC}\, \wedge \, {\mathcal{A}}_{CB} &=&
4 \, {\rm i}  {\Phi}_A^x \, \wedge \, {\Phi}_B^y \, \epsilon_{xy}\nonumber\\
d \Phi^x_A \, +  \, \Delta^{xy} \, \wedge \, \epsilon_{yz} \, \Phi^z_A \, - \,
 e \, {\mathcal{A}}_{AB} \, \wedge \,\Phi^x_B &=&  0
\label{orfan26}
\end{eqnarray}
where all $1$-forms are real and, according to our conventions, the indices $x,y,z,t$ are
symplectic and take four values. The real symmetric bosonic $1$-form $\Omega^{xy} =
\Omega^{yx}$ encodes the generators of the Lie subalgebra
$\sym(4,\mathbb{R})$, while the antisymmetric real bosonic $1$-form
$\mathcal{A}_{AB}= - \mathcal{A}_{BA}$ encodes the generators of
the Lie subalgebra $\so(\mathcal{N})$. The fermionic
$1$-forms $\Phi^x_A$ are real and, as indicated by their
indices, they transform in the fundamental $4$-dim representation of $\sym(4,\mathbb{R})$
and in the fundamental $\mathcal{N}$-dim representation of
$\so(\mathcal{N})$.
Finally,
\begin{equation}
  \epsilon_{xy}= - \epsilon_{yx} \, = \, \left(
  \begin{matrix} 0 & 0 & 0 & 1 \cr 0 & 0 & -1 & 0 \cr 0 & 1 & 0 & 0 \cr
    -1 & 0 & 0 & 0 \cr  \end{matrix} \right)
\label{epsilon}
\end{equation}
is the symplectic invariant metric.
\par
The relation between the formulation (\ref{orfan25}) and
(\ref{orfan26}) of the same Maurer Cartan equations is provided by
the Majorana basis of $d=4$ gamma matrices discussed in appendix
\ref{d4spinorbasis}. Using eq.(\ref{gammareala}), the generators
$\gamma_{ab}$ and $\gamma_a\,\gamma_5$ of the anti de Sitter group
$\mathrm{SO(2,3)}$ turn out to be all given by real symplectic
matrices, as is explicitly shown in eq. (\ref{realgammi}) and the
matrix $\mathcal{C}\,\gamma_5$ turns out to be proportional to
$\epsilon_{xy}$ as shown in eq. (\ref{Chat}). On the other hand a
Majorana spinor in this basis is proportional to a real object times
a phase factor $\exp[ -\, \pi \, {\rm i}\, /\, 4]$.
\par
Hence eq.s (\ref{orfan25}) and eq.s (\ref{orfan26}) are turned ones
into the others upon the identifications:
\begin{equation}
  \begin{array}{ccrcl}
    \Omega^{xy} \, \epsilon_{yz} & \equiv & \Omega^{x}{}_{z} & \leftrightarrow & -  \ft 14 \,
    \omega^{ab} \, \gamma_{ab} \, - \,2\, e \, \gamma_a \, \gamma_5 \, E^a \\
    \null & \null & \mathcal{A}_{AB} & \leftrightarrow & \mathcal{A}_{AB} \\
    \null & \null & \psi_{A}^x & \leftrightarrow & \exp\left[ \ft {- \pi  {\rm i}}{ 4}\right] \, \Phi^x_A \
  \end{array}
\label{conversion}
\end{equation}
As is always the case, the Maurer Cartan equations are just a
property of the (super) Lie algebra and hold true independently of
the (super) manifold on which the $1$-forms are realized: on the supergroup
manifold or on different supercosets of the same supergroup.

\section{The relevant supercosets and their relation}
\label{relevsupcos}

Let us also consider the
following pure fermionic coset:
\begin{eqnarray}
\mathcal{M}^{0|4\mathcal{N}}_{osp} & = & \frac{\mathrm{Osp(\mathcal{N} \, | \, 4
})}{\mathrm{SO(\mathcal{N})} \times \mathrm{Sp(4,\mathbb{R})}}\label{ospsuper0_4n}
\end{eqnarray}
 There is an obvious relation
between these two supercosets that can be formulated in the following
way:
\begin{equation}
   \mathcal{M}^{4|4\mathcal{N}}_{osp} \, \sim \, \mathrm{AdS}_4 \, \times
   \, \mathcal{M}^{0|4\mathcal{N}}_{osp}
\label{relazia}
\end{equation}
In order to explain the actual meaning of eq.(\ref{relazia}) we
proceed as follows. Let the graded matrix $\mathbb{L} \, \in \, \mathrm{Osp(\mathcal{N}|4)}$
be the coset representative of the coset $\mathcal{M}^{4|4 \mathcal{N}}_{osp}$, such that the Maurer Cartan form
$\Lambda$ of eq.(\ref{lambamatra}) can be identified as:
\begin{equation}
  \Lambda = \mathbb{L}^{-1} \, d \mathbb{L}
\label{cosettusrepre}
\end{equation}
Let us now factorize $\mathbb{L}$ as follows:
\begin{equation}
  \mathbb{L} = \mathbb{L}_F \, \mathbb{L}_B
\label{factorL}
\end{equation}
where $\mathbb{L}_F$ is a coset representative for the coset :
\begin{equation}
  \frac{\mathrm{Osp(\mathcal{N} \, | \, 4
})}{\mathrm{SO(\mathcal{N})} \times \mathrm{Sp(4,\mathbb{R})}} \, \ni
\, \mathbb{L}_F
\label{LF}
\end{equation}
and $\mathbb{L}_B$ is the $\mathrm{Osp(\mathcal{N}|4)}$ embedding of a coset
representative of $\mathrm{AdS_4}$, namely:
\begin{equation}
  \mathbb{L}_B \, = \, \left(\begin{array}{c|c}
    \mathrm{L_B} & 0 \\
    \hline
    0 & \mathbf{1}_{\mathcal{N}} \
  \end{array} \right) \quad ; \quad
  \frac{\mathrm{Sp(4,\mathbb{R})}}{\mathrm{SO(1,3)}}\, \ni \,
  \mathrm{L_B}
\label{salamefelino}
\end{equation}
In this way we find:
\begin{equation}
  \Lambda = \mathbb{L}_B^{-1} \, \Lambda_F \, \mathbb{L}_B \, + \, \mathbb{L}_B^{-1}
  \, d \, \mathbb{L}_B
\label{ferrone}
\end{equation}
Let us now write the explicit form of $\Lambda_F $ in analogy to
eq.(\ref{lambamatra}):
\begin{equation}
  \Lambda_F =\left(\begin{array}{c|c}
    \Delta_F & \Theta_A \\
    \hline
    \,  4 \, {\rm i} \, e \, \overline{\Theta}_A \, \gamma_5 & - \, e \, \widetilde{\mathcal{A}}_{AB} \
  \end{array} \right)
\label{fermionform}
\end{equation}
where $\Theta_A$ is a Majorana-spinor valued fermionic $1$-form and
where $\Delta_F$ is an $\sym(4,\mathbb{R})$ Lie algebra valued
$1$-form presented as a $4 \times 4$ matrix. Both $\Theta_A$ as
$\Delta_F$ and $\widetilde{\mathcal{A}}_{AB}$ depend only on the
fermionic $\theta$ coordinates and differentials.
\par
On the other hand we have:
\begin{equation}
  \mathbb{L}_B^{-1}
  \, d \, \mathbb{L}_B \, = \, \left(\begin{array}{c|c}
    \Delta_B & 0 \\
    \hline
    0 & 0 \
  \end{array} \right)
\label{boseforma}
\end{equation}
where the $\Omega_B$ is also an $\sym(4,\mathbb{R})$ Lie algebra valued
$1$-form presented as a $4 \times 4$ matrix, but it depends only on
the bosonic coordinates $x^\mu$ of the anti de Sitter space $\mathrm{AdS_4}$.
Indeed, according to eq(\ref{lambamatra}) we can write:
\begin{equation}
\Delta_B \, = \,    -  \ft 14 \, B^{ab} \, \gamma_{ab} \, - \,2\, e \, \gamma_a \, \gamma_5 \, B^a
\label{Bbwriting}
\end{equation}
where
$\left\{ B^{ab}\, ,\, B^a\right\} $ are respectively the spin-connection and the
vielbein of $\mathrm{AdS_4}$, just as $\left\{ \mathcal{B}^{\alpha\beta}\, ,\, \mathcal{B}^\alpha\right\}
$ are the connection and vielbein of the internal coset manifold
$\mathcal{M}_7$.
\par
Inserting now these results into eq.(\ref{ferrone}) and comparing
with eq.(\ref{lambamatra}) we obtain:
\begin{eqnarray}
\psi_A & = & \mathrm{L_B^{-1} }\, \Theta_A \nonumber\\
\mathcal{A}_{AB} & = & \widetilde{\mathcal{A}}_{AB} \nonumber\\
 -  \frac{1}{4} \, \omega^{ab} \, \gamma_{ab} \, - \,2\, e \, \gamma_a \, \gamma_5 \, E^a
 & = &  -  \frac{1}{4} \, B^{ab} \, \gamma_{ab} \, - \,2\, e \, \gamma_a \, \gamma_5 \,
 B^a \, + \, \mathrm{L_B^{-1} } \, \Delta_F \, \mathrm{L_B }
\label{arcibaldo}
\end{eqnarray}
The above formulae encode an important information. They show how the
supervielbein and the superconnection of the supermanifold
(\ref{ospsuper4_4n}) can be constructed starting from the vielbein
and connection of $\mathrm{AdS_4}$ space plus the Maurer Cartan forms
of the purely fermionic supercoset (\ref{ospsuper0_4n}). In other
words formulae (\ref{arcibaldo}) provide the concrete interpretation
of the direct product (\ref{relazia}). This will also be our starting
point for the actual construction of the supergauge completion in the
case of maximal supersymmetry and for its generalization to the cases
of less supersymmetry.
\subsection{Finite supergroup elements}
We studied the $\osp(\mathcal{N}|4)$
superalgebra but for our purposes we cannot confine ourselves to the
superalgebra, we need also to consider finite elements of the
corresponding supergroup. In particular the supercoset
representative. Elements of the supergroup are described by graded
matrices of the form:
\begin{equation}
  M = \left( \begin{array}{c|c}
    A & \Theta \\
    \hline
   \Pi & D\
  \end{array} \right)
\label{gradamatra}
\end{equation}
where $A,D$ are submatrices made out of even elements of a Grassmann
algebra while $\Theta,\Pi$ are submatrices made out of odd elements
of the same Grassmann algebra. It is important to recall, that the
operations of transposition and hermitian conjugation are defined as
follows on graded matrices:
\begin{eqnarray}
M^T  & = & \left( \begin{array}{c|c}
  A^T & \Pi^T \\
  \hline
 - \,  \Theta^T & D^T
\end{array}\right)  \nonumber\\
M^\dagger  & = & \left( \begin{array}{c|c}
  A^\dagger & \Pi^\dagger \\
  \hline
  \Theta^\dagger & D^\dagger
\end{array} \right)
\label{porfidorosa}
\end{eqnarray}
This is done in order to preserve for the supertrace the same formal
properties enjoyed by the trace of ordinary matrices:
\begin{eqnarray}
\mathrm{Str} \, \left( M\right)  & =  & \mathrm{Tr} \, \left( A\right)  - \mathrm{Tr}\, \left(D\right) \nonumber\\
\mathrm{Str} \, \left( M_1 \, M_2 \right)  & =  & \mathrm{Str} \,
\left(M_2 \, M_1\right)
\label{stracciona}
\end{eqnarray}
Eq.s (\ref{porfidorosa}) and (\ref{stracciona}) have an important
consequence. The consistency of the equation:
\begin{equation}
  M^\dagger = \left( M^T\right) ^\star
\label{starotta}
\end{equation}
implies that the complex conjugate operation on a super matrix must
be defined as follows:
\begin{equation}
  M^\star \, = \, \left( \begin{array}{c|c}
    A^\star & - \Theta^\star \\
    \hline
     \Pi^\star & D^\star \
  \end{array} \right)
\label{staronsupermatra}
\end{equation}
Let us now observe that in the Majorana basis which we have adopted
we have:
\begin{eqnarray}
\widehat{C} & = & {\rm i} \, \, \left(\begin{array}{c|c}
  \epsilon & 0 \\
  \hline
  0 & - \ft 1 {4e} \, \mathbf{1}_{\mathcal{N}\times \mathcal{N}}
\end{array} \right) \, = \, {\rm i} \, \widehat{\epsilon}\nonumber\\
\widehat{H} & =  & \, \left(\begin{array}{c|c}
 {\rm i} \, \epsilon & 0 \\
  \hline
  0 & - \ft 1 {4e} \, \mathbf{1}_{\mathcal{N}\times \mathcal{N}}
\end{array} \right)
\label{porchette}
\end{eqnarray}
where the $4 \times 4$ matrix $\epsilon$ is given by eq.(\ref{Chat}).
Therefore in this basis an orthosymplectic group element $ \mathbb{L} \, \in
\, \OSp(\mathcal{N}|4)$ which satisfies:
\begin{eqnarray}
\mathbb{L}^T \, \widehat{C} \, \mathbb{L}  & = & \widehat{C} \label{reala1}\\
\mathbb{L}^\dagger \, \widehat{H} \, \mathbb{L}  & = & \widehat{H}
\label{reala2}
\end{eqnarray}
has the following structure:
\begin{equation}
  \mathbb{L} \, = \, \left( \begin{array}{c|c}
    \mathcal{S} & \exp \left[-\, {\rm i} \ft {\pi}{4} \right]\Theta \\
    \hline
    \exp \left[-\, {\rm i} \ft {\pi}{4} \right]\,  \Pi & \mathcal{O} \
  \end{array}\right)
\label{struttamatra}
\end{equation}
where the bosonic sub-blocks $\mathcal{S},\mathcal{O}$ are
respectively $4 \times 4 $ and $\mathcal{N}\times \mathcal{N}$ and
real, while the fermionic ones $\Theta,\Pi$ are respectively
$4\times \mathcal{N}$ and $ \mathcal{N}\times 4 $ and also real.
\par
The orthosymplectic conditions (\ref{reala1}) translate into the
following conditions on the sub-blocks:
\begin{eqnarray}
\mathcal{S}^T \, \epsilon \, \mathcal{S} & = &\epsilon - \, {\rm i}\, \ft {1}{4e} \, \Pi^T \, \Pi \nonumber\\
\mathcal{O}^T \,  \mathcal{O} & = & \mathbf{1} \, +  \, {\rm i} \, 4e \,\Theta^T \, \epsilon \,\Theta \nonumber\\
\mathcal{S}^T \, \epsilon \, \Theta & = & -  \, \ft {1}{4e} \, \Pi^T
\, \mathcal {O}
\label{treconde}
\end{eqnarray}
As we see, when the fermionic off-diagonal sub-blocks are zero the
diagonal ones are respectively a symplectic and an orthogonal matrix.
\par
If the graded matrix $\mathbb{L}$ is regarded as the coset
representative of either one of the two supercosets
(\ref{ospsuper4_4n},\ref{ospsuper0_4n}), we can evaluate the explicit
structure of the left-invariant one form $\Lambda$. Using the
$\mathcal{M}^{0|4\times \mathcal{N}}$ style of the Maurer Cartan equations (\ref{orfan26})
we obtain:
\begin{equation}
  \Lambda \, \equiv \, \mathbb{L}^{-1} \, d\mathbb{L} \, = \, \left( \begin{array}{c|c}
    \Delta & \exp \left[- {\rm i } \ft {\pi}{4}\right] \, \Phi \\
    \hline \\
    -   4 e \, \exp \left[ - {\rm i } \ft {\pi}{4}\right] \Phi^T \, \epsilon &  - \, e \, \mathcal{A} \
  \end{array}\right)
\label{ortosymLI}
\end{equation}
where the $1$-forms $\Delta$, $\mathcal{A}$ and $\Phi$ can be
explicitly calculated, using the explicit form of the inverse coset
representative:
\begin{equation}
  \mathbb{L}^{-1} \, = \, \left(\begin{array}{c|c}
    - \epsilon \, \mathcal{S}^T \,  \epsilon & \, \exp \left[- {\rm i } \ft {\pi}{4}\right] \, \ft 1{4e} \,
    \epsilon \Pi^T \, \\
    \hline
   - \exp \left[- {\rm i } \ft {\pi}{4}\right]\, 4e \, \Theta^T \, \epsilon &  \mathcal{O}^T \
  \end{array} \right)
\label{lugubre}
\end{equation}
\begin{eqnarray}
e \mathcal{A} & = & - \, \mathcal{O}^T \, d\mathcal{O} \, - \,{\rm i} \,  4e \, \Theta^T \, \epsilon \, d\Theta \nonumber\\
\Omega  & = & - \, \epsilon \, \mathcal{S}^T \, \epsilon \,
d\mathcal{S}\, - \, {\rm i} \, \ft {1}{4e} \, \Pi^T \, d \Pi \nonumber\\
\Phi & = & \, - \,
\epsilon \, S^T \, \epsilon \, d\Theta \, + \, \ft{1}{4e}
\, \epsilon \, \Pi^T \, d \mathcal{O}
\label{Mcforme}
\end{eqnarray}
\subsection{The coset representative of
$\OSp(\mathcal{N}|4)/\mathrm{SO}(\mathcal{N}) \times \mathrm{Sp(4)}$}
It is fairly simple to write an explicit form for the coset
representative of the fermionic supermanifold
\begin{equation}
 \mathcal{M}^{0|4\times \mathcal{N}} \, = \,  \frac{\OSp(\mathcal{N}|4)}{\mathrm{Sp(4,\mathbb{R})} \times \mathrm{SO}(\mathcal{N})}
\label{supermanifoldamente}
\end{equation}
by adopting the upper left block components $\Theta$ of the
supermatrix (\ref{struttamatra}) as coordinates. It suffices to solve
eq.s(\ref{treconde}) for the sub blocks
$\mathcal{S},\mathcal{O},\Pi$. Such an explicit solution is provided
by setting:
\begin{eqnarray}
\mathcal{O}(\Theta) & = & \left(\mathbf{1} \, + \, 4\,{\rm i}\, e \, \Theta^T \, \epsilon \, \Theta\right) ^{1/2} \nonumber\\
\mathcal{S}(\Theta) & = & \left(\mathbf{1} \, + \, 4\,{\rm i}\, e \, \Theta \,  \Theta^T \, \epsilon \,\right) ^{1/2} \nonumber\\
\Pi & = & {4e}  \left(\mathbf{1} \, + \, 4\,{\rm i}\, e \, \Theta^T \, \epsilon \, \Theta\right) ^{-1/2}
\, \Theta^T \, \epsilon \, \left(\mathbf{1} \, + \, 4\,{\rm i}\, e \, \Theta \,  \Theta^T \, \epsilon \,\right)
^{1/2}\nonumber\\
& = & {4e} \, \Theta^T \, \epsilon \null
\label{explicitone}
\end{eqnarray}
In this way we conclude that the coset representative of the
fermionic supermanifold (\ref{supermanifoldamente}) can be chosen to
be the following supermatrix:
\begin{equation}
  \mathbb{L}\left(\Theta \right)  \, = \, \left( \begin{array}{c|c}
    \left(\mathbf{1} \, + \, 4\,{\rm i}\, e \, \Theta \,  \Theta^T \, \epsilon \,\right) ^{1/2} &
    \exp \left[-\, {\rm i} \ft {\pi}{4} \right]\Theta \\
    \hline
   -\, \exp \left[-\, {\rm i} \ft {\pi}{4} \right]\,  {4e} \, \Theta^T \, \epsilon &
    \left(\mathbf{1} \, + \, 4\,{\rm i}\, e \, \Theta^T \, \epsilon \, \Theta\right) ^{1/2}\
  \end{array}\right)
\label{supercosettus}
\end{equation}
By straightforward steps from eq.(\ref{lugubre}) we obtain the inverse of the  supercoset
element (\ref{supercosettus}) in the form:
\begin{equation}
  \mathbb{L}^{-1}\,\left(\Theta \right)  \, = \, \mathbb{L}\,\left(- \, \Theta \right) \, = \, \left( \begin{array}{c|c}
    \left(\mathbf{1} \, + \, 4\,{\rm i}\, e \, \Theta \,  \Theta^T \, \epsilon \,\right) ^{1/2} &
   - \, \exp \left[-\, {\rm i} \ft {\pi}{4} \right]\Theta \\
    \hline
    \exp \left[-\, {\rm i} \ft {\pi}{4} \right]\,  {4e} \, \Theta^T \, \epsilon &
    \left(\mathbf{1} \, + \, 4\,{\rm i}\, e \, \Theta^T \, \epsilon \, \Theta\right) ^{1/2}\
  \end{array}\right)
\label{supercosettusminus}
\end{equation}
Correspondingly we work out the explicit expression  of the Maurer Cartan forms:
\begin{eqnarray}
e \mathcal{A} & = &  \,  \left(\mathbf{1} \, + \, 4\,{\rm i}\, e \, \Theta^{T} \,  \epsilon \,  \Theta \right) ^{1/2} \, d
 \left(\mathbf{1} \, + \, 4\,{\rm i}\, e \,  \Theta^T \, \epsilon \, \Theta \, \right) ^{1/2}
 \, - \,{\rm i} \,  4e \, \Theta^T \, \epsilon \, d\Theta \nonumber\\
\Phi & = &  \left(\mathbf{1} \, + \, 4\,{\rm i}\, e \,  \Theta \,  \Theta^T \,\epsilon \right) ^{1/2} \,  d\Theta \, + \, \Theta \,
d \left(\mathbf{1} \, + \, 4\,{\rm i}\, e \,  \Theta^T \, \epsilon \, \Theta
\right)^{1/2} \nonumber\\
\Delta & = & \left(\mathbf{1} \, + \, 4\,{\rm i}\, e \,  \Theta \,
\Theta^T \epsilon \,
\right)^{1/2} \, d \, \left(\mathbf{1} \, + \, 4\,{\rm i}\, e \,  \Theta \, \Theta^T \epsilon \,
\right)^{1/2} \, - \, {\rm i} \, 4e \, \Theta \, d\Theta^T \, \epsilon
\label{Mcforme2}
\end{eqnarray}

 \section{Osp pure spinors}

Having discussed the properties of the supergroup and its cosets, we develop
the technique of "ghostyfying" the MC forms. This was already discussed in \cite{marp,Fre:2008qw}
and it amounts to extending the differential $d$ entering the definition of the MC equations to
a BRST differential and to extending the fermionic MC forms with a ghost field $\Lambda$. The
latter is a bosonic variable which will be identified with the pure spinor variable.

We first fermionize the MC
forms for ${\mathrm{Osp}(\mathcal{N} \, | \, 4})$ and we derive the set of pure spinor conditions for a generic $\mathcal{N}$.
Then we compare this set
of constraints with the constraints found from the
supergravity approach and we discuss the  number of independent parameters.
Next, we consider
the case of those supercosets that are of the form
${\mathrm{Osp}(\mathcal{N} \, | \, 4})/ \mathrm{SO}(\mathcal{N} -1) \times \mathrm{SO(1,3)}$. Then, we consider
the cases ${\mathrm{Osp}(\mathcal{N} \, | \, 4})/ U(\mathcal{N}/2) \times \mathrm{SO(1,3)}$ where
$\mathcal{N}$ is an even number, and finally the case of the fermionic Grassmannian
${\mathrm{Osp}(\mathcal{N} \, | \, 4})/ \mathrm{SO}(\mathcal{N} ) \times \mathrm{Sp(4,\mathbb{R})}$.
These cases produce three different types of pure spinor constraints that we analyze.

\subsection{PS for ${\mathrm{Osp}(\mathcal{N} \, | \, 4})$ }

We recall the Maurer Cartan equations (\ref{orfan26})
\begin{eqnarray}
d \Delta^{xy} + \Delta^{xz} \, \wedge \,\Delta^{ty} \, \epsilon_{zt} &=&
 -\, 4 \, {\rm i}\,  e \,  {\Phi}_A^x \, \wedge \, {\Phi}_A^y, \nonumber \\
d {\mathcal{A}}_{AB} - e  {\mathcal{A}}_{AC}\, \wedge \, {\mathcal{A}}_{CB} &=&
4 \, {\rm i}  {\Phi}_A^x \, \wedge \, {\Phi}_B^y \, \epsilon_{xy}\nonumber\\
d \Phi^x_A \, +  \, \Delta^{xy} \, \wedge \, \epsilon_{yz} \, \Phi^z_A \, - \,
 e \, {\mathcal{A}}_{AB} \, \wedge \,\Phi^x_B &=&  0\,,
\label{newMC}
\end{eqnarray}
and we fermionize them by substituting $ d \rightarrow d + {\mathcal S}$ and
$\Phi^x_A \rightarrow \Phi^x_A + \Lambda^x_A$. In addition, we do not add any ghost field to the
bosonic MC forms. This is equivalent to say that we are not gauging
any subgroup of the supergroup, but we are gauging only the fermionic variables.
This interpretation is not completely satisfactory and we refer to \cite{Grassi:2004cz} for a
more detailed discussion.

This yields the transformations rules
\begin{eqnarray}
 {s}\,  \Delta^{xy} &=&
 -\, 4 \, {\rm i}\,  e \,  {\Lambda}_A^{(x} \, {\Phi}_B^{y)} \delta^{AB} \,, \nonumber \\
 {s}\,  {\mathcal{A}}_{AB}  &=&\,
4 \, {\rm i}  \, {\Lambda}_{[A}^x \, \, {\Phi}_{B]}^y \, \epsilon_{xy} \,,\nonumber\\
 {s} \, \Phi^x_A \,  &=& - \, d \Lambda^x_A \, -
 \, \Delta^{xy} \,  \epsilon_{yz} \, \Lambda^z_A \, + \,
 e \, {\mathcal{A}}_{AB}  \,\Lambda^x_C \, \delta^{BC} \,, \nonumber \\
 {s}\,  \Lambda^x_A &=& 0 \,.
\label{newMC_fer}
\end{eqnarray}
and the pure spinor constrains
\begin{eqnarray}
{\Lambda}_A^{(x} \, {\Lambda}_B^{y)} \, \delta^{AB} \, =0 \,, \hspace{2cm}
{\Lambda}_{[A}^{x} \, {\Lambda}_{B]}^{y}  \, \epsilon_{xy}  \,  =0 \,.
\label{newMC_ps}
\end{eqnarray}
The BRST transformations for $\Delta^{xy}, \mathcal{A}_{AB}$ and $\Phi^x_A$ are nilpotent.
This follows from the pure spinor constraints (\ref{newMC_ps}) and
from the (anti)symmetrization of the spinorial indices of $\Lambda^x_A$. Notice that
we have traded the form degree with the ghost number passing from $\Phi^x_A$ to $\Lambda^x_A$.

This set of constraints are not all independent. Indeed, by contracting the first equation
with $\Lambda^z_C \epsilon_{yz}$, because of the second equation, it automatically
vanishes. In the same way, by hitting the second equation with $\Lambda^z_C  \delta^{CA}$,
we get a redundant equation.

Now, suppose that we solve the first set of equations, the matrix
$\mathcal{G}_{[AB]} = {\Lambda}_{[A}^{x} \, {\Lambda}_{B]}^{y}  \, \epsilon_{xy}$ is
antisymmetric and also nilpotent. However, any vector
of the form $\Lambda^x_C \delta^{AC} F_x$ (with $\Lambda^x_A$ solution of the first set of equations)
is an eigenvector of $\mathcal{G}_{[AB]}$. This means that, if there are enough parameters in the
solution $\Lambda^x_A$, such that there are enough independent eigenvectors
$\Lambda^x_C \delta^{AC} F_x$, then the matrix $\mathcal{G}_{[AB]}$ should vanish. However,
it can be checked that there are solutions of the first equation which do not satisfy the second equation.

One convenient way to parametrize the solution is to decompose the $\mathrm{Sp(4,\mathbb{R})}$ index into $\mathrm{SO(1,3})$
irreducible representations. Since the vector representation of $\mathrm{Sp(4,\mathbb{R})}$ is isomorphic to
the spinorial representation of $\mathrm{SO(1,3)}$ we use the latter and we decompose the indices
$x,y,\dots$ into $\alpha, \dot\alpha, \beta, \dot\beta,\dots$. Then eqs. (\ref{newMC_ps}) can rewritten
as follows:
  \begin{eqnarray}
&&{\Lambda}_A^{(\alpha} \, {\Lambda}_B^{\beta)} \, \delta^{AB} \, =0 \,, \hspace{2cm}
{\Lambda}_A^{(\dot\alpha} \, {\Lambda}_B^{\dot\beta)} \, \delta^{AB} \, =0 \,, \hspace{2cm}
{\Lambda}_A^{(\alpha} \, {\Lambda}_B^{\dot\beta)} \, \delta^{AB} \, =0 \,, \nonumber \\
&&\hspace{3.5cm}
{\Lambda}_{[A}^{\alpha} \, {\Lambda}_{B]}^{\beta}  \, \epsilon_{\alpha\beta} +
{\Lambda}_{[A}^{\dot\alpha} \, {\Lambda}_{B]}^{\dot\beta}  \, \epsilon_{\dot\alpha\dot\beta}
\,  =0 \,.
\label{newMC_psB}
\end{eqnarray}
We decompose the pure spinors $\lambda^\alpha_A$ and $\lambda^{\dot \alpha}_A$
in the factorized form
\begin{equation}\label{factA}
\lambda^\alpha_A = \lambda^\alpha v_A\,,
\hspace{2cm}
\lambda^{\dot\alpha}_A = \lambda^{\dot \alpha} u_A\,, ~~~~
\end{equation}
where $\lambda^\alpha$ and $\lambda^{\dot\alpha}$ are two spinors of $\mathrm{SO(1,3)}$, while
$u_A$ and $v_A$ are vectors of $\mathrm{SO(\mathcal{N})}$. Notice that the decomposition (\ref{factA}) implies two independent gauge symmetries
 $\lambda^\alpha \rightarrow \rho \lambda^\alpha$ and
$\lambda^{\dot\alpha} \rightarrow \sigma \lambda^{\dot\alpha}$ which are compensated by
the transformations of $u_A \rightarrow \sigma^{-1} u_A$ and $v_A \rightarrow \rho^{-1} v_A$.  Inserting
factorization (\ref{factA}) in eqs. (\ref{newMC_psB}) yields the following remaining constraints
\begin{equation}\label{factB}
u_A u_B \, \delta^{AB} =0\,, \hspace{2cm}
u_A v_B \, \delta^{AB} =0\,,\hspace{2cm}
v_A v_B \, \delta^{AB} =0\,,
\end{equation}
which can be easily solved by adopting a light-cone decomposition of vectors $u_A$ and $v_A$.
Let us count the parameters: we get $2 \times (2 + \mathcal{N} -1)$ from decomposition (\ref{factA})
(the $-1$ comes from the gauge symmetries) and we impose the scalar constraints (\ref{factB}). This leads to $2\mathcal{N} -1$
parameters in the solution.

If we sum the bosonic coordinates $10 + \mathcal{N}(\mathcal{N}-1)/2$ (associated
with
the bosonic subgroup) to the pure spinors $2\mathcal{N} -1$ minus the
fermionic coordinates $4 \mathcal{N}$, we find that there is no solution with the match of the bosonic
and fermionic degrees of freedom. Even though,
it seems consistent to construct a pure spinor model associated with the ${\mathrm{Osp}(\mathcal{N} \, | \, 4})$
supergroup manifold, we do not have a string theory interpretation.

\subsection{PS for ${\mathrm{Osp}(\mathcal{N} \, | \, 4})/ \mathrm{SO}(\mathcal{N} -1) \times \mathrm{SO(1,3)}$}

As a second example, we consider the coset
${\mathrm{Osp}(\mathcal{N} \, | \, 4})/ \mathrm{SO}(\mathcal{N} -1) \times \mathrm{SO(1,3)}$ where we gauge the
subgroup $\mathrm{SO}(\mathcal{N} -1) \times \mathrm{SO(1,3)}$ of the bosonic subgroup
$\mathrm{SO}(\mathcal{N}) \times \mathrm{Sp(4,\mathbb{R})}$. From the
supergravity point of view this would correspond a compactification
on a background of the form $\mathrm{AdS_4} \, \times
\mathbb{S}^{\mathcal{N}}$.

Technically, our choice means that  we add the ghost fields $\xi_{IJ}$ associated with the
subgroup $\mathrm{SO}(\mathcal{N} -1)$ (where $I,J=1,\dots,\mathcal{N} -1$) and the ghost field $\xi_{ab}$ (where $a,b=1,\dots, 4$) associated with
the Lorentz group $\mathrm{SO(1,3)}$.
For that, we decompose the matrix $\Delta^{xy} = \Delta^a \gamma^{xy}_a +
\Delta^{ab} \gamma^{xy}_{ab}$ and the matrix ${\mathcal A}_{AB} = ({\mathcal A}_I,{\mathcal A}_{IJ})$.
Accordingly, we decompose the MC equations. However, since we have now introduced the ghost
fields associated with the MC forms $ \Delta^{ab}$ and ${\mathcal A}_{IJ}$, we can reabsorb the
non-vanishing right-hand side of MC equations by the BRST transformations of the new ghost fields
except for the "pure spinor" constraints
\begin{equation}
{\Lambda}_A^{x} \, \gamma^a_{xy }\, {\Lambda}_B^{y} \, \delta^{AB} \, =0 \,, \hspace{2cm}
{\Lambda}_{I}^{x} \, {\Lambda}^{y}  \, \epsilon_{xy}  \,  =0 \,.
\label{newppss}
\end{equation}
where we have decomposed the fermionic MC form $\Phi^x_A$
into  $(\Phi^x, \Phi^x_I)$. For ${\mathcal N} =8$, we can use  triality to relate the vector index to spinor index and
rewrite the second constraint as ${\Lambda}^{x} \, \tau^\alpha\, {\Lambda}^{y}  \, \epsilon_{xy}  \,  =0$
where $\alpha =1,\dots, 7$.

The relevant BRST transformations are
\begin{eqnarray}
 {s}\,  \Delta^{a} + \xi^{ab} \Delta^b &=&
 -\, 4 \, {\rm i}\,  e \,  {\Lambda}_A^{x}\, \gamma^a_{xy} \, {\Phi}_B^{y} \,\delta^{AB} \,, \nonumber \\
 {s}\,  {\mathcal{A}}_{I} + \xi_{IJ}\, {\mathcal A}_J &=&\,
4 \, {\rm i}  \, {\Lambda}_{I}^x \, \, {\Phi}^y \, \epsilon_{xy} \,,\nonumber\\
 {s} \, \Phi^x_A \,  + \gamma_{ab}^{xy} \, \xi^{ab} \, \Phi^y_A + \delta_{A K} \, {\xi}_{KI} \, \Phi^x_I
 &=& - \, d \Lambda^x_A \, -
 \, \Delta^{xy} \,  \epsilon_{yz} \, \Lambda^z_A \, + \,
 e \, {\mathcal{A}}_{AB}  \,\Lambda^x_C \, \delta^{BC} \,, \nonumber \\
 {s}\,  \Lambda^x_A
 + \gamma_{ab}^{xy} \, \xi^{ab} \, \Lambda^y_A + \delta_{A K} \, {\xi}_{KI} \, \Lambda^x_I &=& 0 \,, \nonumber \\
 {s}\, \xi^{ab} + \xi^{ac} \xi_c^{~b} &=& {\Lambda}_A^{x} \,
 \gamma^{ab}_{xy}\, {\Lambda}_B^{y} \, \delta^{AB} \,, \nonumber  \\
 {s}\, \xi_{IJ} + \xi_{IK} \xi^K_{~J} &=&
{\Lambda}_{[I}^{x} \, {\Lambda}^{y}_{J]}  \, \epsilon_{xy} \,.
 \label{newMC_ferB}
\end{eqnarray}
which are nilpotent because of the pure spinor constraints (\ref{newppss}).

In addition, one can define a "covariant" BRST differential $s_\xi$ by reabsorbing the
ghosts $\xi_{ab}$ and $\xi_{IJ}$. Then we can rewrite the first three expressions
in (\ref{newMC_ferB}) as follows
\begin{eqnarray}
 {s_\xi}\,  \Delta^{a}  &=&
 -\, 4 \, {\rm i}\,  e \,  {\Lambda}_A^{x}\, \gamma^a_{xy} \, {\Phi}_B^{y} \,\delta^{AB} \,, \nonumber \\
 {s_\xi}\,  {\mathcal{A}}_{I} &=&\,
4 \, {\rm i}  \, {\Lambda}_{I}^x \, \, {\Phi}^y \, \epsilon_{xy} \,,\nonumber\\
 {s_\xi} \, \Phi^x_A \,
 &=& - \, d \Lambda^x_A \, -
 \, \Delta^{xy} \,  \epsilon_{yz} \, \Lambda^z_A \, + \,
 e \, {\mathcal{A}}_{AB}  \,\Lambda^x_C \, \delta^{BC} \,,
 \label{newMC_ferC}
\end{eqnarray}
which look similar to the orginal transformations. An important note:
the new fields $\xi_{IJ}$ and $\xi^{ab}$ are not dynamical fields and they are just needed
in order to make the gauge invariance manifest. The corresponding sigma model must be
gauge invariant under the symmetries of the subgroup and therefore the new ghost fields
do not enter the action. If the $\xi$'s were to be dynamical, we would have to take them into account
for counting the degrees of freedom.

Again, we can count the number of independent parameters in the pure spinor constraints. We can notice that in the case of maximal
supersymetry ($\mathrm{SO}(8)$) the two set of constraints reproduce
the 11 dimensional pure spinor constraints. However, for lower dimension the counting has to be performed.
We adopt the same decomposition for the pure spinors $\Lambda^x_A$ as for the fermionic MC
forms $\Phi^x_A$ and we use the $\mathrm{SO(1,3)}$ spinorial indices $\alpha, \dot\alpha$ for simplicity.
Eqs. (\ref{newppss}) are re-written as follows
\begin{eqnarray}
{\Lambda}_I^{(\alpha} \, {\Lambda}_J^{\dot\beta)} \, \delta^{IJ} +
{\Lambda}^{(\alpha} \, {\Lambda}^{\dot\beta)} \, =0 \,, \hspace{2cm}
{\Lambda}_{I}^{\alpha} \, {\Lambda}^{\beta}  \, \epsilon_{\alpha\beta} +
{\Lambda}_{I}^{\dot\alpha} \, {\Lambda}^{\dot\beta}  \, \epsilon_{\dot\alpha\dot\beta}
\,  =0 \,.
\label{newMC_psBib}
\end{eqnarray}

Then, we propose the ansatz
\begin{equation}\label{ansA}
\Lambda^\alpha_I = \lambda^\alpha u_I\,, \hspace{1cm}
\Lambda^{\dot\alpha}_I = \lambda^{\dot\alpha} v_I\,, \hspace{1cm}
\Lambda^\alpha = \lambda^\alpha \,, \hspace{1cm}
\Lambda^{\dot\alpha} = \lambda^\alpha \,,
\end{equation}
which inserted in (\ref{newMC_psBib}) leads to the remaining constraint
\begin{equation}\label{newpp}
u_I v_J \delta^{IJ} + 1  =0\,.
\end{equation}
Then, counting the contraints and the dof, we get that the number
of independent parameters for the pure spinors (\ref{newMC_psBib}) is
$2\mathcal{N} +1$. Notice that there is no gauge symmetry left in the present case since
 $\Lambda^\alpha$ and $\Lambda^{\dot\alpha}$ are not gauge invariant.

Summing the bosonic coordinates $4 + (\mathcal{N}-1)$ (notice that the internal space is a sphere
$\mathrm{SO(\mathcal{N})} / \mathrm{SO(\mathcal{N}-1)}$, the pure spinor coordinates $2 \mathcal{N} +1$ minus the fermionic coordinates
$4 \mathcal{N}$ we get a single solution for $\mathcal{N}=4$. This is a remarkable result since the
coset $\mathrm{Osp(4|4)}/ \mathrm{SO(1,3)} \times \mathrm{SO(3)}$ corresponds a bosonic background $\mathrm{AdS_4} \times \mathbb{S}^3$
which is a background for $d=7$ supergravity compactified on a 3-sphere. It could be understood
as the compatification of 11d supergravity on $\mathbb{P}^2$ (this breaks the supersymmetry
from $\mathcal{N} =8$ down to $\mathcal{N} =4$) leading to a d=7 supergravity with such an amount of supersymmetry..

\subsection{PS for  ${\mathrm{Osp}(\mathcal{N} \, | \, 4})/\mathrm{ U(\mathcal{N}/2)} \times \mathrm{SO(1,3)}$}

The coset ${\mathrm{Osp}(\mathcal{N} \, | \, 4})/ \mathrm{SO(}\mathcal{N} -1) \times \mathrm{SO(1,3)}$ is not
the only interesting possibility. For example, for $\mathcal{N} = 2 n$, we can
divide by the maximal subgroup $\mathrm{U(n)}$. This means that we have to add the ghost fields
associated with the generators of the subgroup $\mathrm{U(n)}$ and therefore we have to decompose the
generators $T_{[AB]}$ of $\mathrm{SO(\mathcal{N})}$ according to irreducible
representations of the chosen subgroup as follows $(T_{[IJ]}, T_I^{~\bar J}, T^{[\bar I\bar J]})$.
The generators of the subgroup are identified with $T_I^{~\bar J}$
and we associate the ghosts $\xi^I_{~\bar J}$ to them. Thus, the pure spinor constraints
become
 \begin{equation}
{\Lambda}_A^{x} \, \gamma^a_{xy }\, {\Lambda}_B^{y} \, \delta^{AB} \, =0 \,, ~~~~~~~~~
{\Lambda}_{[I}^{x} \, {\Lambda}^{y}_{J]}  \, \epsilon_{xy}  \,  =0 ~~~~~~~
{\Lambda}_{~[\bar I}^{x} \, {\Lambda}^{y}_{{\bar J}]}  \, \epsilon_{xy}  \,  =0 \,.
\label{newppssB}
\end{equation}
There are 4 constraints for the $\mathrm{Sp(4,\mathbb{R})}$ part, and $n (n-1)$ constraints for the internal part to be compared with
$2 n$ constraints in (\ref{newppssB}).
For example in the case of ${\mathcal N} =6$, we have  $4 + 3  + 3 = 10$ constraints. This case
has a supercoset of the form ${\mathrm{Osp}(6\, | \, 4})/ \mathrm{U(3)} \times \mathrm{SO(1,3)}$ which is
the appropriate supergroup for the $\mathrm{AdS_4} \times {\mathbb P}^3$ supergravity solution.

To solve the pure spinor constraints (\ref{newppssB}), we use again the decomposition into
$\mathrm{SO(1,3)}$ spinorial indices and we decompose the index $A$ into $I$ and $\bar I$ with
$I,\bar I=1, \dots,n$. The constraints read
\begin{eqnarray}
&& {\Lambda}_I^{(\alpha} \, {\Lambda}_J^{\dot\beta)} \, \delta^{IJ} +
{\Lambda}_{\bar I}^{(\alpha} \, {\Lambda}_{\bar J}^{\dot\beta)} \delta^{\bar I \bar J}\, =0 \,, \hspace{2cm}
\nonumber \\
&&{\Lambda}_{[I}^{\alpha} \, {\Lambda}_{J]}^{\beta}  \, \epsilon_{\alpha\beta} +
{\Lambda}_{[I}^{\dot\alpha} \, {\Lambda}_{J]}^{\dot\beta}  \, \epsilon_{\dot\alpha\dot\beta}
\,  =0 \,, \\
&&{\Lambda}_{[\bar I}^{\alpha} \, {\Lambda}_{\bar J]}^{\beta}  \, \epsilon_{\alpha\beta} +
{\Lambda}_{[\bar I}^{\dot\alpha} \, {\Lambda}_{\bar J]}^{\dot\beta}  \, \epsilon_{\dot\alpha\dot\beta}
\,  =0 \,, \nonumber
\label{newMC_psBibA}
\end{eqnarray}
To solve them, we use the factorization
\begin{eqnarray}\label{factCC}
&&\Lambda^\alpha_I = \lambda^\alpha u_I\,, \hspace{3cm}
\Lambda^{\dot\alpha}_I = \lambda^{\dot \alpha} v_I\,, \\
&&
\Lambda^\alpha_{\bar I} = \bar \lambda^\alpha u_{\bar I}\,, \hspace{3cm}
\bar \Lambda^{\dot\alpha}_{\bar I} = \bar\lambda^{\dot \alpha} v_{\bar I}\,, \nonumber
 \end{eqnarray}
defined up to $\mathbb{C}^*$ gauge transformations
$$\lambda^\alpha \rightarrow \sigma \lambda^\alpha, \hspace{2cm}
 \lambda^{\dot \alpha} \rightarrow \rho \lambda^{\dot \alpha}\,, \hspace{2cm}
\bar\lambda^\alpha \rightarrow \bar\sigma \bar\lambda^\alpha, \hspace{2cm}
\bar\lambda^{\dot \alpha} \rightarrow \bar\rho \bar\lambda^{\dot \alpha}\,.$$
and analogously for $v_I, u_I, v_{\bar I}$ and $u_{\bar I}$.

 Inserting these factorizations into (\ref{newMC_psBibA}), we arrive at the constraints
 \begin{equation}\label{factCD}
 u_I v_J \delta^{IJ} =0\,, \hspace{3cm}  u_{\bar I} v_{\bar J} \delta^{\bar I\bar J} =0\,.
 \end{equation}
So, computing the number of independent degrees of freedom we get
$4 \times (2 + \mathcal{N}/2 -1) - 2= 2 \mathcal{N} + 2$. For $\mathcal{N}=6$ we get exactly 14 pure spinors.
In addition, by summing the bosonic coordinates $4 + \mathcal{N}(\mathcal{N}-1)/2 - \mathcal{N}^2/4$ and the
pure spinor contribution $2 \mathcal{N} + 2$ minus the fermionic coordinates   $4 \mathcal{N}$. It turns
out that there are only two solutions with the matching of the bosonic and fermionic degrees of
freedom for $\mathcal{N}=6$ and $\mathcal{N}=4$. The first case corresponds to the background
$\mathrm{Osp}(6|4)/ \mathrm{U(3)} \times \mathrm{SO(1,3)}$, the latter to
the background $\mathrm{Osp}(4|4)/ \mathrm{U(2)} \times \mathrm{SO(1,3)}$. The first one is a
background for the critical type IIA d=10 superstring with the bosonic background $\mathrm{AdS_4} \times \mathbb{P}^3$
and with the RR forms $\mathbb{G}^{[2]} \propto \mathcal{K}$ (where $\mathcal{K}$ is the  K\"alher 2-form on $\mathbb{P}^3$)
and $\mathbb{G}^{[4]} = e \mathrm{Vol}_{4}$ where $\mathrm{Vol}_{4}$ is the volume form of the $\mathrm{AdS_4}$-space.
The supersymmetry is $\mathcal{N}=6$
for a specific choice of the dilaton v.e.v.. This example is the $\mathbb{S}^1$ reduction of the round $\mathbb{S}^7 \times \mathrm{AdS_4}$
solution of 11d which has originally $\mathcal{N}=8$ supersymmetries and looses two of them in the reduction.
The catch of the reduction is the Hopf fibration of the round seven sphere: $\mathbb{S}^7 \stackrel{\pi}{\Longrightarrow}
\mathbb{P}^3$.
The second example corresponds to a  non-critical d=6 superstring
(or better for d=6 supergravity) with the bosonic background $\mathrm{AdS_4} \times \mathbb{P}^1$ and the
RR forms  $\mathbb{G}^{[2]} \propto \mathcal{K}$ (where $\mathcal{K}$ is the  K\"alher 2-form on $\mathbb{P}^13$)
and $\mathbb{G}^{[4]} = e \mathrm{Vol}_{4}$ where $\mathrm{Vol}_{4}$ is the volume form of the $\mathrm{AdS_4}$-space. Notice the also
 $\mathbb{S}^3$ has a Hopf fibration: $\mathbb{S}^3 \stackrel{\pi}{\Longrightarrow}
\mathbb{P}^1 $ so that we can argue that this model can be obtained from the $\mathbb{S}^3 \times
\mathrm{AdS_4}$ supergravity background discussed above.
 The residual supersymmetry is $\mathcal{N}=4$.

\subsection{PS  for ${\mathrm{Osp}(\mathcal{N} \, | \, 4})/ \mathrm{SO(\mathcal{N})} \times \mathrm{Sp(4,\mathbb{R})}$}

By dividing the supgroup by the entire bosonic subgroup, we mean that we add the complete set of ghost
fields associated with the generators of $\mathrm{SO(\mathcal{N})}$ and of $\mathrm{Sp(4,\mathbb{R})}$. This means that
all MC forms have their own ghost extension and therefore there is no pure spinor constraint
left. Notice that in this case we have for any $\mathcal{N}$ a complete matching between
the pure spinor fields and fermionic fields. This situation is described as a gauged linear sigma
model by Berkovits and Vafa in \cite{Berkovits:2007rj}. The sigma model can be constructed as a WZW model and
the corresponding Ka\v c-Moody algebra realizes the loop generalization of the algebra of the coset.

\section{Pure Spinor Sigma Model for $\mathrm{AdS}_4 \times \mathbb{CP}^3$}

The sigma model can be decomposed in the following pieces: 
\begin{equation}
\mathcal{S} = \mathcal{S}_1 + \mathcal{S}_2 + \mathcal{S}_3 + \mathcal{S}_4
\end{equation}
where 
\begin{equation}\label{kinA}
\mathcal{S}_1 = \int e^+\wedge e^- (\eta^{ab} J_{a +} J_{b -} + 
\, J_{I J,  +} J^{IJ}_{-}  + 
\, J_{I J,  -} J^{IJ}_{+} ) 
\end{equation}
in the conformal gauge. 
To make contact with the standard notation in the literature on sigma models, we introduce new names 
for the pull-back on the worldsheet of the MC forms 
$(X: \Sigma_2 \rightarrow \mathrm{AdS}_4 \times \mathbb{P}^3)$
$$
X^*(\Delta)= - \frac{1}{4} \gamma_{ab} \, H^{ab} - 2 \, e\, (\gamma^a \gamma^5) J_a\,, 
$$
$$
X^*( \mathcal{A}_{AB}) =   U^{~~JK}_{[AB]} \, J_{JK} +  H_{I }^{~J} U^I_{~J, AB}  +U_{[AB] IJ} \, J^{J K}  \,,
$$
and $ U_{IJ}^{~~AB}, \dots,  U_{I}^{~ J,AB}$ are the matrices converting the 
$\mathrm{SO}(6)$ vector representation into $U(3)$ basis. So, $J_{IJ}, J^{I J}$ are the MC forms  
associated to generators of the coset ${\rm SU}(4)/{\rm U}(3)$ and $H_I^{~J}$ are the MC of the 
generators of ${\rm U}(3)$. $\eta^{ab}$ is the invariant metric on ${\rm AdS}_4$ and $g_{I\bar J}$ 
is the ${\rm U}(3)$ invariant metric, we denote by $k_{I \bar J}$ the components of the K\"alher form on $\mathbb{P}^3$. 
The index $I$ can be raised and lowered with the metric $g^{I\bar J}$, for example $J^{\bar I \bar J} = g^{\bar I K} \, 
g^{\bar J L}\, J_{KL}$ which is independent of $J^{IJ}$. 

The MC equations discussed in (\ref{orfan26}) can be rewritten using the complex coordinates. We 
separate the $H$-connections $H^{ab}, H_I^{~J}$ from the vielbeins $J_a, J_{IJ}, J^{IJ}, \Phi_I, \Phi^I$
\begin{eqnarray}\label{MCa}
 R^{ab} &\equiv& d H^{ab} - H^{ac} \wedge H^{db} \, \eta_{cd} = - 16 \, e^2\, J^a \wedge J^b  - 2 \, e \, \overline \Phi_I \wedge \gamma^{ab} \gamma^5 \Phi^I \,, \nonumber \\
 R_I^{~J} &\equiv& d H_I^{~~J} -  H_I^{~~K} \wedge H_K^{~~J} =  e^2 \, J_{I K} \wedge J^{K J} +
4 \, e\, \overline\Phi_I \wedge\, \gamma^5  \, \Phi^J   \,, \nonumber \\
 \nabla J^a &\equiv& d J^a - H^a_{~b} \wedge J^b = \frac{1}{2} \, 
 \overline \Phi_I \wedge \gamma^a \Phi^I\,, \nonumber \\
\nabla J_{IJ} &\equiv&  d J_{IJ} - 2\, H_{[I}^{~K} \wedge \, J_{J] K} =
4\, \overline\Phi_I \wedge\, \gamma^5  \, \Phi_J \,,
\nonumber \\
\nabla \Phi_I &\equiv& d \Phi_I - \frac{1}{4} H^{ab} \wedge \gamma_{ab} \Phi_I  -   H_I^{~J} \wedge \Phi_J
 = e \, J_{IJ} \wedge \Phi^J + 2 \, e\, J^a \wedge \gamma_a \gamma_5 \Phi_I\,.
  \end{eqnarray}
  The MC equations for $J^{IJ}$ and $\Phi^I$ are obtained by conjugation from the last twos.
  The covariant derivatives are taken with respect to the gauge group $\mathrm{U(3) \times SO(1,3)}$.
  It is also convenient to adopt another basis by decomposing the spinorial indices $x,y,z,...$ into
 $\mathrm{SO(1,3)}$ indices. In particular, we decompose the spinorial MC forms
 $\Phi^x_I$ and $\Phi^{x I}$ as follows $\Phi^\alpha_I, \Phi^{\dot \alpha}_I$ and $\Phi^{\alpha I}, \Phi^{\dot \alpha I}$.
 Now, grouping these spinors into the two
 sets $(\Phi^\alpha_I, \Phi^{\dot\alpha}_{\bar I})$ (where $\Phi^{\dot\alpha}_{\bar I} = g_{\bar I J} \Phi^{\dot \alpha J}$)
 and $(\Phi^{\dot \alpha \bar I},\Phi^{\alpha I})$, (where
 $\Phi^{\dot \alpha \bar I}= g^{\bar I J} \Phi^{\dot \alpha}_I$) we can organize the MC forms
 into the following subsets:
 \begin{eqnarray}\label{MCz4}
{\mathcal H}_0 &=& \Big\{ H^{ab}, H_I^{~J} \Big\}\,,  \hspace{.5cm}
{\mathcal H}_1 = \Big\{ \Phi^\alpha_I,  \Phi^{\dot\alpha}_{\bar I} \Big\}\,, \nonumber \\
{\mathcal H}_2 &=& \Big\{ J^a, J_{IJ}, \overline{J}^{IJ} \Big\}\,, ~~~~~~~
{\mathcal H}_3 = \Big\{\Phi^{\alpha I}, \Phi^{\dot\alpha \bar I}  \Big\}\,.
\end{eqnarray}
and so doing the $\mathrm{Osp}(6|4)$ algebra acquires a $\mathbb{Z}_4$ grading (as it is   the case of $\mathrm{PSU(2,2|4)}$).
This discrete symmetry is very useful for deriving the non-local conserved charges \cite{Bena:2003wd}.
 Notice that we have derived it for the MC forms, but it can be obviously discussed
 at the level of the algebra. Again, there is an overlap between our results and the results in \cite{Arutyunov:2008if,Stefanski:2008ik}.

One of the important features of the supergravity background we are discussing is the possibility
to write the Wess-Zumino term as a total derivative of a globally defined quantity. It reads as follows
\begin{eqnarray}\label{wzA}
\mathcal{H} &=& 4  \,e\, J_a  \wedge \overline \Phi_I \wedge\, (\gamma^a\gamma^5)  \Phi^{J} +
e \, J_{IJ}\,  \wedge \overline\Phi^I \wedge  \Phi^J  +
e\, \overline J^{I J} \wedge  \overline \Phi_{I} \wedge \Phi_{J}  \\
&=& d\,  \Big( 2\, {\rm i} \, \overline \Phi_{I} \, \wedge \Phi^{I} \Big)\,.
\end{eqnarray}
and therefore we can write it on the 2d surface as
\begin{eqnarray}\label{wzB}
\mathcal{S}_2 = 2\, {\rm i} \alpha \, \int \overline \Phi_{I} \, \wedge \Phi^{I}  =  2 {\rm i}   \alpha\,
\int e^+\wedge e^- \, (\overline\Phi_{I +} \, \Phi^I_{-} -
\overline\Phi_{I -} \, \Phi^I_{+}  ) \,.
\end{eqnarray}
where we have introduced a constant $\alpha$ in front of the WZ term. Notice that the WZ term is written
by means
of $\mathrm{SO(1,3)}$ and $\mathrm{U(3)}$ invariant tensors. The constant $\alpha$ is fixed by $\kappa$-symmetry
which can be easily derived from the MC forms. In particular, we derive the general variation under a fermionic shift
$\Phi_I, \Phi^I \rightarrow \Phi_I + \epsilon_I, \Phi^I+ \epsilon^I$ where $\epsilon^I, \epsilon_I$ are commuting spinors.
(in previous sections we have denoted them by $\Lambda_I$ and $\Lambda^I$ and we have derived the pure spinor conditions).
Then we have the variations
 \begin{eqnarray}\label{MCa}
&& \delta J^a =  \frac{1}{2}\, \overline \epsilon_I  \gamma^a \Phi^I +  \frac{1}{2}
\, \overline \Phi_I  \gamma^a \epsilon^I\,, \nonumber \\
&& \delta J_{IJ} =4 \, \overline\Phi_{[I} \gamma^5  \, \epsilon_{J]} \,,
\nonumber \\
&& \delta \Phi_I = e \, J_{IJ} \epsilon^J + 2 \, e\, J^a (\gamma_a \gamma_5 \epsilon)_I\,.
  \end{eqnarray}
It turns out that the action ($\mathcal{S}_1+\mathcal{S}_2$) is invariant if $\alpha = 1/(4\, e)$ and if the
spinors $\epsilon_I, \epsilon^I$ satisfy a suitable projection. This is similar to the
$\kappa$-transformation of the $\mathrm{AdS_5}\times \mathbb{S}^5$ model and we find that the there is a relation between the worldsheet
chirality, the target space chirality and the K\"alher structure of $\mathbb{P}^3$, as expected. It can be proved that the $\kappa$-symmetry
reduces consistently to 16 coordinates (which can be chosen to be light-cone coordinates). We refer to
papers \cite{Arutyunov:2008if,Stefanski:2008ik} for a discussion on this point since we are interested in the pure spinor
construction.\footnote{We recall that the sigma model for plane-wave has been constructed and discussed in
\cite{Sugiyama:2002tf,Nishioka:2008gz,Gaiotto:2008cg,Grignani:2008te}.}

So, the Green-Schwarz action (in the conformal gauge) is given by the simple quadratic action
\begin{equation}\label{totA}
\mathcal{S}_1 = \int e^+\wedge e^- \left( \eta^{ab} J_{a +} J_{b -} + 
\, J_{I J,  +} J^{IJ}_{-}  + 
\, J_{I J,  -} J^{IJ}_{+}  + \frac{ i }{2 \, e} (\overline\Phi_{I +} \, \Phi^I_{ -} -
\overline\Phi_{I -} \, \Phi^I_{+}) \right)
\end{equation}
written in term of the MC forms. The coupling constant can be put as an overall constant by redefining the MC forms.
In order to see the discrete symmetry manifestly, we can rewrite the WZ term as follows
$$
\int e^+\wedge e^- (\Phi_{I \alpha +} \Phi^{I \alpha}_- -
\Phi_{I \alpha -} \Phi^{I \alpha}_+
 - \Phi_{I \dot\alpha +} \Phi^{I \dot\alpha}_-   + \Phi_{I \dot\alpha -} \Phi^{I \dot\alpha}_+  
 )
$$ 
which
has the structure of $\mathcal{H}_1 \times \mathcal{H}_3$ with respect to $\mathbb{Z}_4$ discrete symmetry.

The third term contains the RR fields $\mathbb{G}^{[4]}$ and $\mathbb{G}^{[2]}$. We recall that the
4d RR field is of the form $\mathbb{G}^{[4]} = e \epsilon_{abcd} J^a\wedge \dots \wedge J^d$ and
$\mathbb{G}^{[2]} = k_{I \bar J} J^I \wedge J^{\bar J}$ where $k_{I\bar J}$ is the K\"alher form
on $\mathbb{P}^3$ and $J^I = \epsilon^{IJK} J_{JK}$ and $J^{\bar J} = g^{\bar J J} \epsilon_{JKL} J^{KL}$. 
In the case of the $\mathrm{AdS_5}\times \mathbb{S}^5$ background and in the case of
non-critical superstrings (see \cite{Adam:2007ws}), the form of the RR term is unique. Namely, due to
the isometries, the form of the term is fixed. In the present case the invariance under
$\mathrm{U(3) \times SO(1,3)}$ is not sufficient to fix completely the RR terms and one requires the BRST
symmetry to do it. In a parallel work we find a systematic way to produce the correct RR couplings
\cite{marp2}.

As is been mentioned, we should add some new additional fields
associated to the pure spinor setting. We introduce the conjugated momenta
$d_{I z},d_{z}^I$ and the anti-holomorphic ones $d_{I \bar z},d_{\bar z}^I$. The form
of the action is \cite{berko-pp}
\begin{eqnarray}\label{RRA}
\mathcal{S}_3 &=& \int e^+\wedge e^- \Big(
 \overline d_{+}\,  ({\bf 1}_4 \otimes {\bf 1}_6 +  i \gamma_5 \otimes k_6) \Phi_- +
\overline d_{-}\,  ({\bf 1}_4 \otimes {\bf 1}_6 -  i \gamma_5 \otimes k_6) \Phi_+  \nonumber \\
 &+& i\, e \, \overline d_{+}
\Big({\bf 1}_4 \otimes k_6  - 3\, i \, \gamma_5 \otimes {\bf 1}_6 \Big) d_{-} \Big)
\end{eqnarray}
where we recall that $e$ is the coupling constant and it represents the flux of the RR
field. The form of the matrix between the two $d$'s  has been derived using the formalism \cite{marp}, 
and a complete derivation will be presented elsewhere \cite{marp2}. 
Since the $d$-terms can be integrated we get a simplified action
\begin{eqnarray}\label{RRB}
\mathcal{S}_3 
 &=&
 - \frac{i}{4 \, e} \int e^+\wedge e^- (\Phi_{I \alpha+ } \Phi^{I \alpha}_- + \Phi^I_{\dot\alpha +} \Phi_{I -}^{\dot\alpha})
\end{eqnarray}
The last term of the action contains two invariants, namely $1 \otimes g_{I \bar J}$ and
$\gamma_5 \otimes k_{I \bar J}$ which are made of invariants under $\mathrm{SO(1,3) \times U(3})$
and the linear combination of these two invariants appearing in the action is fixed by the BRST symmetry. Notice that, differently from
the case of $\mathrm{PSU}(2,2|4)$, there are two invariants and this might imply that the model
is not conformal invariant. However, this must be checked by an explicit one-loop computation.
Nevertheless, it seems that the form of the RR-term reproduces the cases known in the literature
\cite{Berkovits:2004xu} and \cite{Adam:2007ws} where the WZ term combines in a non-trivial way with the
RR term producing a kinetic term for the fermions which is no longer invariant under $\kappa$-symmetry and
therefore can be quantized.

We introduce the pure spinor Lorentz generators which are needed in the action and
they determine the couplings between the pure spinor fields and the matter fields. In addition, they
give the coupling with the Riemann tensor.
\begin{eqnarray}\label{LG_A}
N^{ab}_L &=& 
\frac{1}{2} \overline w^I  \gamma^{ab} (1 + \gamma^5) \lambda_I +
\frac{1}{2} \overline w_I  \gamma^{ab} (1 + \gamma^5) \lambda^I 
\,, \hspace{1cm} \\
N^{ab}_R &=& 
\frac{1}{2} \overline w^I  \gamma^{ab} (1 - \gamma^5) \lambda_I +
\frac{1}{2} \overline w_I  \gamma^{ab} (1 - \gamma^5) \lambda^I 
\,, \hspace{1cm}
\\
N_{I}^{~J} &=& \frac{1}{2} \overline w_{I} \lambda^J\,, \hspace{1.2cm}
\bar N_{I}^{~J} = \frac{1}{2} \overline w^J \lambda_{I}\,, ~~~~
 \end{eqnarray}
 The overline stands for the Dirac coniugation and they are gauge invariant under the
 gauge transformations  generated by the pure spinor constraints
 \begin{eqnarray}\label{loppo}
\delta w^I = \Xi_a (\g^a \l)^I + \Gamma^{IJ} (\gamma^5 \l)_J\,, \hspace{1cm}
\delta w_I = \Xi_a (\g^a \l)_I + \Gamma_{IJ} (\gamma^5 \l)^J\,,
\end{eqnarray}
where $\Xi_a, \Gamma_{IJ}$ and $\Gamma^{IJ}$ are the gauge parameters of the
infinitesimal transformations.
It is also convenient to write them in the spinorial notation to get the two combinations of the first two operators
\begin{eqnarray}\label{lollo}
&&N_{\alpha\beta} = w^I_{(\alpha} \lambda_{\beta)I} + w_{I (\alpha} \lambda_{\beta)}^I\,,
\hspace{1cm}
N_{\dot\alpha\dot\beta} = w^I_{(\dot\alpha} \lambda_{\dot\beta)I} + w_{I (\dot\alpha} \lambda_{\dot\beta)}^I\,, \nonumber \\
&&
N_I^{~J} =  w^I_\a \lambda^\a_I +  w_{I \dot\a} \lambda^{I \dot\a}\,,
\hspace{1.2cm}
\bar N_I^{~J} =  w_I^\a \lambda_\a^I +  w^{I \dot\a} \lambda_{I \dot\a}\,.
\end{eqnarray}
Finally, in terms of these ingredients, we can write the last piece of the action
\begin{equation}\label{S4}
\mathcal{S}_4 = \int e^+\wedge e^- \Big( \overline w_{I +} \nabla_- \l^I + \overline w^I_{-} \overline\nabla_+ \l_I +
R_{ab, cd} N^{ab}_{+} N^{cd}_- +
R^{I~~K}_{~J, ~~ L} \, N_{I +}^{~J} \, \bar N_{K -}^{~L} \Big)
\end{equation}
where $R_{ab, cd}$ is the Riemann tensor of the $\mathrm{AdS_4}$ space and
$R^{I~~K}_{~J, ~~ L} $ is the Riemann tensor of the internal space $\mathbb{P}^3$.
To check that all the pieces of the action fit together, we need to impose the BRST symmetry. This
can be done by constructing BRST variations:
Then we have the variations
 \begin{eqnarray}\label{Brst}
&& {\cal S} J^a =  \frac{1}{2}\, \overline \lambda_I  \gamma^a \Phi^I +  \frac{1}{2}
\, \overline \Phi_I  \gamma^a \lambda^I\,, \nonumber \\
&& {\cal S} J_{IJ} =4 \, \overline\Phi_{[I} \gamma^5  \, \lambda_{J]} \,,
\nonumber \\
&& {\cal S} \Phi_I = \nabla \lambda_I + e \, J_{IJ} \lambda^J + 2 \, e\, J^a (\gamma_a \gamma_5 \lambda)_I\,.
  \end{eqnarray}
The BRST charge is nilpotent because of the pure spinor constraints and due to the gauge invariance under the gauge group
$\mathrm{U(3) \times SO(1,3)}$ and the invariance of the action can be checked by acting with the BRST charge on the different pieces of the action.
We do not write here the computation since the structure of the action and of the BRST charge looks very similar
to the one presented in \cite{Berkovits:2004xu,Adam:2007ws} and therefore it can be analyzed by the same steps.
Furthermore in the shortly forthcoming paper \cite{marp2} we show
that the action described in the present article can be exactly
derived by localizing on the $\mathrm{AdS_4} \times \mathbb{P}^3$
background the action discussed in \cite{marp} which was shown there
to be BRST invariant on a generic supergravity background.

The supersymmetry $\mathrm{N}=6$ preserved by the background is still quite strong to imply the equations of motion, therefore
we expect that the BRST charges applied to a generic vertex operator imply that the background fields are on-shell. In any case,
this point deserves further investigations since we know examples such as those described in \cite{Adam:2006bt} where this does not happen.

\section{Conclusions and Future Work}

We have discussed several examples of $\mathrm{AdS_4}$ backgrounds viewed as coset spaces of the
supergroup $\mathrm{Osp}(\mathcal{N}|4)$. We analyzed the pure spinor constraints in all cases and we found that
only few of them admit an interpretation as supergravity backgrounds. Moreover, we discussed in detail the case of
$\mathrm{AdS_4} \times \mathbb{P}^3$ and we wrote down the Green-Schwarz model and the corresponding pure spinor action. The
latter is more convenient since it has all 24 supersymmetries manifest. Notice, as was discovered in \cite{Berkovits:2004xu} the
supersymmetry invariance of the action does not require any boundary term in contrast to the flat case. In addition, one can perform
the limit as in \cite{Berkovits:2007zk} and the model can be described in terms of a gauged linear sigma model based
on the superGrassmannian space $\mathrm{Osp}(6|4)/ \mathrm{SO(6)} \times \mathrm{Sp(4,\mathbb{R})}$. It would be very interesting to see
what the amplitudes compute in the present context and we have to study the corresponding measure. We notice that
as in the $\mathrm{AdS_5} \times \mathbb{S}^5$ case, there are singleton representations and it would be interesting to see whether
 one of these singleton
representations of $\mathrm{AdS_4}$ reduces to a superconformal Chern-Simons theory on the boundary in analogy with the AdS/CFT duality for
$\mathrm{AdS_5} \times \mathbb{S}^5$ and for $\mathrm{AdS_4} \times \mathbb{S}^7$ \cite{Ferrara:1997dh,Dall'Agata:1998wz}.

In a forthcoming paper \cite{marp2}, we analyze the pure spinor sigma model from the geometric perspective using
the construction in \cite{marp}. In that context the pure spinor constrains can be derived from the rheonomic parametrization
of  type IIA supergravity. In order to adapt the rheonomic parametrization to the case $\mathrm{AdS_4} \times \mathbb{P}^3$
we specify all terms in the action given in \cite{marp}.

\section*{Acknowledgments}

We are grateful to R. D'Auria, M. Trigiante, G. Dall'Agata, D. Sorokin, and M. Tonin for useful discussions. P.A.G. is grateful
to P. Vanhove for invitation at the Institute for Theoretical Physics, Saclay, Paris where part of this work has been completed. P.A.G. would like to thank R. Roiban for valuable comments. 

 \newpage
\appendix
\section{D=6 gamma matrix basis}
\label{d7spinorbasis}
In the discussion of the $\mathrm{AdS_4} \times \mathbb{P}^3$
compactification we need to consider the decomposition of
the $d=10$ gamma matrix algebra into the tensor product of the
$\so(6)$ clifford algebra times that of $\so(1,3)$.
In this section we discuss and explicit basis for the $\so(6)$
gamma matrix algebra using that of $\so(7)$. Conventionally we
identify the $7$-matrix $\tau_7$ with the chirality matrix in $d=6$.
\par
In this paper, the indices $\alpha,\beta,\dots$ run on six values
and denote the vector indices of $\so(6)$. In order to discuss the
gamma matrix basis we introduce $\so(7)$ indices
\begin{equation}
  \overline{\alpha} = \alpha, 7
\label{extendedalpha}
\end{equation}
which run on seven values and we define
the Clifford algebra with negative metric:
\begin{equation}
  \left\{ \tau_{\overline{\alpha}} \, , \, \tau_{\overline{\beta}}\right\}
   \, = \, - \delta_{\overline{\alpha \beta}}
\label{taualgebra}
\end{equation}
This algebra is satisfied by the following, real, antisymmetric matrices:
{\scriptsize
\begin{eqnarray*}
  \begin{array}{ccccccc}
    \tau_1 & = & \left(
\begin{array}{llllllll}
 0 & 0 & 0 & 0 & 0 & 0 & 0 & 1 \\
 0 & 0 & 1 & 0 & 0 & 0 & 0 & 0 \\
 0 & -1 & 0 & 0 & 0 & 0 & 0 & 0 \\
 0 & 0 & 0 & 0 & 0 & 0 & 1 & 0 \\
 0 & 0 & 0 & 0 & 0 & -1 & 0 & 0 \\
 0 & 0 & 0 & 0 & 1 & 0 & 0 & 0 \\
 0 & 0 & 0 & -1 & 0 & 0 & 0 & 0 \\
 -1 & 0 & 0 & 0 & 0 & 0 & 0 & 0
\end{array}
\right) &; & \tau_2 & = & \left(
\begin{array}{llllllll}
 0 & 0 & -1 & 0 & 0 & 0 & 0 & 0 \\
 0 & 0 & 0 & 0 & 0 & 0 & 0 & 1 \\
 1 & 0 & 0 & 0 & 0 & 0 & 0 & 0 \\
 0 & 0 & 0 & 0 & 0 & 1 & 0 & 0 \\
 0 & 0 & 0 & 0 & 0 & 0 & 1 & 0 \\
 0 & 0 & 0 & -1 & 0 & 0 & 0 & 0 \\
 0 & 0 & 0 & 0 & -1 & 0 & 0 & 0 \\
 0 & -1 & 0 & 0 & 0 & 0 & 0 & 0
\end{array}
\right) \\
\end{array}
\end{eqnarray*}
\begin{eqnarray*}
  \begin{array}{ccccccc}
    \tau_3 & = & \left(
\begin{array}{llllllll}
 0 & 1 & 0 & 0 & 0 & 0 & 0 & 0 \\
 -1 & 0 & 0 & 0 & 0 & 0 & 0 & 0 \\
 0 & 0 & 0 & 0 & 0 & 0 & 0 & 1 \\
 0 & 0 & 0 & 0 & -1 & 0 & 0 & 0 \\
 0 & 0 & 0 & 1 & 0 & 0 & 0 & 0 \\
 0 & 0 & 0 & 0 & 0 & 0 & 1 & 0 \\
 0 & 0 & 0 & 0 & 0 & -1 & 0 & 0 \\
 0 & 0 & -1 & 0 & 0 & 0 & 0 & 0
\end{array}
\right) & ; & \tau_4 & = & \left(
\begin{array}{llllllll}
 0 & 0 & 0 & 0 & 0 & 0 & -1 & 0 \\
 0 & 0 & 0 & 0 & 0 & -1 & 0 & 0 \\
 0 & 0 & 0 & 0 & 1 & 0 & 0 & 0 \\
 0 & 0 & 0 & 0 & 0 & 0 & 0 & 1 \\
 0 & 0 & -1 & 0 & 0 & 0 & 0 & 0 \\
 0 & 1 & 0 & 0 & 0 & 0 & 0 & 0 \\
 1 & 0 & 0 & 0 & 0 & 0 & 0 & 0 \\
 0 & 0 & 0 & -1 & 0 & 0 & 0 & 0
\end{array}
\right)  \
\end{array}
\end{eqnarray*}
\begin{eqnarray*}
  \begin{array}{ccccccc}
   \tau_5 & = & \left(
\begin{array}{llllllll}
 0 & 0 & 0 & 0 & 0 & 1 & 0 & 0 \\
 0 & 0 & 0 & 0 & 0 & 0 & -1 & 0 \\
 0 & 0 & 0 & -1 & 0 & 0 & 0 & 0 \\
 0 & 0 & 1 & 0 & 0 & 0 & 0 & 0 \\
 0 & 0 & 0 & 0 & 0 & 0 & 0 & 1 \\
 -1 & 0 & 0 & 0 & 0 & 0 & 0 & 0 \\
 0 & 1 & 0 & 0 & 0 & 0 & 0 & 0 \\
 0 & 0 & 0 & 0 & -1 & 0 & 0 & 0
\end{array}
\right) & ; & \tau_6 & = & \left(
\begin{array}{llllllll}
 0 & 0 & 0 & 0 & -1 & 0 & 0 & 0 \\
 0 & 0 & 0 & 1 & 0 & 0 & 0 & 0 \\
 0 & 0 & 0 & 0 & 0 & 0 & -1 & 0 \\
 0 & -1 & 0 & 0 & 0 & 0 & 0 & 0 \\
 1 & 0 & 0 & 0 & 0 & 0 & 0 & 0 \\
 0 & 0 & 0 & 0 & 0 & 0 & 0 & 1 \\
 0 & 0 & 1 & 0 & 0 & 0 & 0 & 0 \\
 0 & 0 & 0 & 0 & 0 & -1 & 0 & 0
\end{array}
\right)  \
\end{array}
\end{eqnarray*}
\begin{equation}
  \begin{array}{ccc}
    \tau_7 & = & \left(
\begin{array}{llllllll}
 0 & 0 & 0 & 1 & 0 & 0 & 0 & 0 \\
 0 & 0 & 0 & 0 & 1 & 0 & 0 & 0 \\
 0 & 0 & 0 & 0 & 0 & 1 & 0 & 0 \\
 -1 & 0 & 0 & 0 & 0 & 0 & 0 & 0 \\
 0 & -1 & 0 & 0 & 0 & 0 & 0 & 0 \\
 0 & 0 & -1 & 0 & 0 & 0 & 0 & 0 \\
 0 & 0 & 0 & 0 & 0 & 0 & 0 & 1 \\
 0 & 0 & 0 & 0 & 0 & 0 & -1 & 0
\end{array}
\right)  \
  \end{array}
\label{tauexplicit}
\end{equation}
}\subsection{D=4 $\gamma$-matrix basis and spinor identities}
\label{d4spinorbasis}
In this section we construct a basis of $\so(1,3)$ gamma matrices
such that it explicitly realizes the isomorphism $\so(2,3)  \sim
\sym (4,\mathbb{R})$ with the conventions used in the main text.
Naming $\sigma_i$ the standard Pauli matrices:
\begin{equation}
  \sigma_1 =\left( \begin{array}{cc}
    0 & 1 \\
    1 & 0 \
  \end{array}\right) \quad ; \quad \sigma_2 =\left( \begin{array}{cc}
    0 & -{\rm i} \\
    {\rm i} & 0 \
  \end{array}\right) \quad ; \quad \sigma_3 =\left( \begin{array}{cc}
    1 & 0 \\
    0 & -1 \
  \end{array}\right)
\label{paulini}
\end{equation}
we realize the $\so(1,3)$ Clifford algebra:
\begin{equation}
  \left\{ \gamma_a \, , \, \gamma_b \right\} \, = \, 2 \, \eta_{ab}
  \quad ; \quad \eta_{ab} \, = \, \mbox{diag} \left( + , - , - , - \right)
\label{d4clif}
\end{equation}
by setting:
\begin{equation}
  \begin{array}{ccccccc}
    \gamma_0 & = &  \sigma_2 \, \otimes \, \mathbf{1}  & ; & \gamma_1 & = & {\rm i} \, \sigma_3  \,
    \otimes \, \sigma_1\\
    \gamma_2 & = & {\rm i} \sigma_1 \, \otimes \, \mathbf{1} & ; &
    \gamma_3 & = & {\rm i} \sigma_3 \, \otimes \, \sigma_3 \\
    \gamma_5 & = & \sigma_3 \, \otimes \, \sigma_2 & ; & \mathcal{C} & = & {\rm i} \sigma_2 \,
    \otimes \, \mathbf{1}
  \end{array}
\label{gammareala}
\end{equation}
where $\gamma_5$ is the chirality matrix and $\mathcal{C}$ is the
charge conjugation matrix. Making now reference to eq.s
(\ref{OmandHmat}) and (\ref{ortosymp}) of the main text we see that
the antisymmetric matrix entering the definition of the
orthosymplectic algebra, namely $\mathcal{C}\, \gamma_5$ is the
following one:
\begin{equation}
  \mathcal{C}\, = \, {\rm i} \left(
  \begin{matrix} 0 & 0 & 1 & 0 \cr 0 & 0 & 0 & 1 \cr -1 & 0 & 0 & 0 \cr
    0 & -1 & 0 & 0 \cr  \end{matrix} \right)\,, \hspace{2cm}
  \mathcal{C}\, \gamma_5 \, = \epsilon = \, {\rm i} \left(
  \begin{matrix} 0 & 0 & 0 & 1 \cr 0 & 0 & -1 & 0 \cr 0 & 1 & 0 & 0 \cr
    -1 & 0 & 0 & 0 \cr  \end{matrix} \right)
\label{Chat}
\end{equation}
namely it is proportional, through an overall ${\rm i}$-factor, to a real
completely off-diagonal matrix. On the other hand all the generators
of the $\so(2,3)$ Lie algebra, \textit{i.e.} $\gamma_{ab}$ and
$\gamma_a\, \gamma_5$ are real, symplectic $4 \times 4$ matrices.
Indeed we have
\begin{equation}
  \begin{array}{ccccccc}
    \gamma_{01} & = &\left(  \begin{matrix} 0 & 0 & 0 & -1 \cr 0 & 0 & -1 & 0 \cr 0 &
    -1 & 0 & 0 \cr -1 & 0 & 0 & 0 \cr  \end{matrix}\right)  & ; & \gamma_{02} & = & \left(
     \begin{matrix} 1 & 0 & 0 & 0 \cr 0 & 1 & 0 & 0 \cr 0 & 0 &
    -1 & 0 \cr 0 & 0 & 0 & -1 \cr  \end{matrix}\right) \\
    \null & \null & \null & \null & \null & \null & \null \\
    \gamma_{12} & = &\left(  \begin{matrix} 0 & 0 & -1 & 0 \cr 0 & 0 & 0 & 1 \cr
    -1 & 0 & 0 & 0 \cr 0 & 1 & 0 & 0 \cr  \end{matrix}\right) & ; & \gamma_{13} & = &\left(  \begin{matrix} 0 & 0 & 0 & -1 \cr 0 & 0 &
    -1 & 0 \cr 0 & 1 & 0 & 0 \cr 1 & 0 & 0 & 0 \cr  \end{matrix}\right)\\
    \null & \null & \null & \null & \null & \null & \null \\
    \gamma_{23} & = & \left(  \begin{matrix} 0 & 1 & 0 & 0 \cr
    -1 & 0 & 0 & 0 \cr 0 & 0 & 0 & 1 \cr 0 & 0 & -1 & 0 \cr  \end{matrix}\right) & ; & \gamma_{34} & = &\left(
     \begin{matrix} 0 & 0 & 1 & 0 \cr 0 & 0 & 0 & -1 \cr
    -1 & 0 & 0 & 0 \cr 0 & 1 & 0 & 0 \cr  \end{matrix}\right) \\
    \null & \null & \null & \null & \null & \null & \null \\
    \gamma_{0} \,\gamma_5 & = & \left(  \begin{matrix} 0 & 0 & 0 & 1 \cr 0 & 0 & -1 & 0 \cr 0 & 1 & 0 & 0 \cr
    -1 & 0 & 0 & 0 \cr  \end{matrix}\right) & ; & \gamma_{1} \,\gamma_5 & = &\left(
    \begin{matrix} -1 & 0 & 0 & 0 \cr 0 & 1 & 0 & 0 \cr 0 & 0 &
    -1 & 0 \cr 0 & 0 & 0 & 1 \cr  \end{matrix}\right) \\
    \null & \null & \null & \null & \null & \null & \null \\
    \gamma_{2} \,\gamma_5 & = & \left(  \begin{matrix} 0 & 0 & 0 & -1 \cr 0 & 0 & 1 & 0 \cr 0 & 1 & 0 & 0 \cr
    -1 & 0 & 0 & 0 \cr  \end{matrix}\right)& ; & \gamma_{3} \,\gamma_5 & = &\left(  \begin{matrix}
   0 & 1 & 0 & 0 \cr 1 & 0 & 0 & 0 \cr 0 & 0 & 0 & 1 \cr 0 & 0 &
   1 & 0 \cr  \end{matrix}\right) \
  \end{array}
\label{realgammi}
\end{equation}
On the other hand we find that $\mathcal{C}\gamma_0 = {\rm i} \,
\mathbf{1}$. Hence the Majorana condition becomes:
\begin{equation}
  {\rm i} \, \psi \, = \, \psi^\star
\label{Majorana}
\end{equation}
so that a Majorana spinor is just a real spinor multiplied by an
overall phase $\exp \left [- i \frac \pi 4\right ]$.
\par
These conventions being fixed let $\chi_x$ ($x=1,\dots,4$) be a set of
(commuting) Majorana spinors normalized in the following way:
\begin{equation}
  \begin{array}{lclcl}
    \chi_x & = & \mathcal{C}\, \overline{\chi}_x^T & ; & \mbox{Majorana condition} \\
    \overline{\chi}_x \, \gamma_5 \, \chi_y & = & {\rm i} \, \left( \mathcal{C}\, \gamma_5\right) _{xy} & ; &
    \mbox{symplectic normal basis} \
  \end{array}
\label{festoso}
\end{equation}
Then by explicit evaluation we can verify the following Fierz
identity:
\begin{equation}
  \ft 12 \, \gamma^{ab} \, \chi_z \, \overline{\chi}_x \, \gamma_5 \,
  \gamma_{ab} \, \chi_y \, - \, \gamma_a \, \gamma_5 \, \chi_z \,
  \overline{\chi}_x \, \gamma_a \, \chi_y \, = \, - \, 2{\rm i} \, \left[
  \left( C\gamma_5\right) _{zx} \, \chi_y \, + \, \left( C\gamma_5\right) _{zy} \,
  \chi_x \right]
\label{firzusd4x1}
\end{equation}
Another identity which we can prove by direct evaluation is the
following one:
\begin{eqnarray}
& \overline{\chi}_x \, \gamma_5 \gamma_{ab} \, \chi_y \, \overline{\chi}_z \, \gamma^b \, \chi_t
\, - \, \overline{\chi}_z \, \gamma_5 \gamma_{ab} \, \chi_t \, \overline{\chi}_x \, \gamma^b \, \chi_y
= &\nonumber\\
&  {\rm i}\left( \overline{\chi}_x \, \gamma_a \, \chi_t \, \left(  \mathcal{C}\,
\gamma_5\right)_{yz} \, + \, \overline{\chi}_y \, \gamma_a \, \chi_t \, \left( \mathcal{C}\,
\gamma_5\right)_{xz}  + \overline{\chi}_x \, \gamma_a \, \chi_z \, \left( \mathcal{C}\,
\gamma_5\right)_{yt} \, + \, \overline{\chi}_y \, \gamma_a \, \chi_z \, \left( \mathcal{C}\,
\gamma_5\right)_{xt}\right) &\nonumber\\
\label{firzusd4x2}
\end{eqnarray}

\par
Finally let us mention some relevant formulae for the derivation of
the compactification. With the above conventions we find:
\begin{equation}
  \gamma_0\,\gamma_1 \, \gamma_2\, \gamma_3 \, =  \, {\rm i} \,
  \gamma_5
\label{g5ide}
\end{equation}
and if we fix the convention:
\begin{equation}
  \epsilon_{0123} \, = \, + \, 1
\label{conveps}
\end{equation}
we obtain:
\begin{equation}
  \ft {1}{24} \, \epsilon^{abcd} \, \gamma_a \, \gamma_b \, \gamma_c
  \, \gamma_d \, = \, - \, {\rm i} \, \gamma_5
\label{epsgamma}
\end{equation}


\end{document}